\documentclass[useAMS,usenatbib,usegraphicx]{mn2e}

\usepackage{dcolumn}
\usepackage[bookmarks,ps2pdf,breaklinks=true]{hyperref}

\hypersetup{citecolor={black},linkcolor={black},urlcolor={black}}  

\hypersetup{
  colorlinks=true,             
  bookmarksnumbered=true,      
  bookmarksopen=false,         
  pdfpagemode=UseOutlines,     
  pdfdisplaydoctitle,          
  pdftitle={PINGS: the PPAK IFS Nearby Galaxies Survey}
}

\usepackage[all]{hypcap}

\newcommand{\hi}{H\,{\footnotesize I}~}
\newcommand{\hh}{H\,{\footnotesize II}~}
\newcommand{\nii}{[N\,{\footnotesize II}]~}
\newcommand{\sii}{[S\,{\footnotesize II}]~}
\newcommand{\oi}{[O\,{\footnotesize I}]~}
\newcommand{\oii}{[O\,{\footnotesize II}]~}
\newcommand{\oiii}{[O\,{\footnotesize III}]~}
\newcommand{\lam}{$\lambda$}

\defcitealias{paperII}{Paper~II}
\defcitealias{Sanchez:2006p331}{San06}
\defcitealias{Moustakas:2006p307}{MK06}
\defcitealias{Ferguson:1998p224}{FGW98}
\defcitealias{Pastoriza:1993p3323}{Pas93}

\title[PINGS: the PPAK IFS Nearby Galaxies Survey]{PINGS: the PPAK IFS Nearby
  Galaxies Survey\thanks{Based on observations collected at the Centro
    Astron\'omico Hispano Alem\'an (CAHA) at Calar Alto, operated jointly by
  the Max-Planck Institut f\"ur Astronomie and the Instituto de Astrof\'isica
  de Andaluc\'ia (CSIC).}
}
\author[Rosales-Ortega et al.]
{F.~F.~Rosales-Ortega,$^{1}$\thanks{E-mail: frosales@ast.cam.ac.uk}
R.~C.~Kennicutt,$^{1}$  S.~F.~S{\'a}nchez,$^{2,3,4}$ A.~I.~D{\'i}az,$^{5}$
\newauthor A.~Pasquali,$^{6}$ B.~D.~Johnson$^{1}$ and C.~N.~Hao$^{1}$\\ 
$^{1}$Institute of Astronomy, University of Cambridge, Madingley Road, Cambridge
CB3 0HA, UK\\
$^{2}$Centro Astron{\'o}mico Hispano Alem{\'a}n, Calar Alto, CSIC-MPG, Jes{\'u}s
Durb{\'a}n Rem{\'o}n 2, E-04004 Almeria, Spain\\
$^{3}$Centro de Estudios de F{\'i}sica del Cosmos de Aragon (CEFCA),
C/General Pizarro 1, 3$^{\circ}$, E-41001 Teruel, Spain\\
$^{4}$Fundaci{\'o}n Agencia Aragonesa para la Investigaci{\'o}n y el
Desarrollo (ARAID)\\
$^{5}$Departamento de F{\'i}sica Te{\'o}rica, C-XI, Universidad Aut{\'o}noma de
Madrid, 28049 Madrid, Spain\\
$^{6}$Max-Planck Institut f{\"u}r Astronomie, K{\"o}nigstuhl 17, 69117
Heidelberg, Germany
}

\begin{document}

\date{Accepted 2010 February 09. Received 2010 January 24; in original form
  2009 May 11}

\pagerange{\pageref{firstpage}--\pageref{lastpage}} \pubyear{2010}

\maketitle

\label{firstpage}

\begin{abstract}
We present the PPAK Integral Field Spectroscopy (IFS) Nearby Galaxies Survey:
PINGS, a 2-dimensional spectroscopic mosaicking of 17 nearby
disk galaxies in the optical wavelength range. This project represents the
first attempt to obtain continuous coverage spectra of the whole surface of a
galaxy in the nearby universe. The final data set comprises more than 50000
individual spectra, covering in total an observed area of nearly 80 arcmin$^2$. 
The observations will be supplemented with broad band and narrow band
imaging for those objects without public available images in order to
maximise the scientific and archival value of the data set.
In this paper we describe the main astrophysical issues to be addressed by the
PINGS project, we present the galaxy sample and explain the observing
strategy, the data reduction process and all uncertainties involved.
Additionally, we give some scientific highlights extracted from the
first analysis of the PINGS sample. A companion paper will report on the first
results obtained for NGC\,628: the largest IFS survey on a single galaxy.
\end{abstract}


\section{Introduction}
\label{sec:intro}

Hitherto, most spectroscopic studies in nearby galaxies have focused on the
derivation of physical and chemical properties of spatially-resolved bright
individual \hh regions.
Most of these measurements were made with
single-aperture or long-slit spectrographs, resulting in samples of typically
a few \hh regions per galaxy or single spectra of large samples like the
Sloan Digital Sky Survey \citep[SDSS,][]{York:2000p2677} or other large surveys.
The advent of multi-object and integral field spectrometers with large fields
of view now offers us the opportunity to undertake a new generation of 
emission-line surveys, based on samples of scores to hundreds of individual
\hh regions within a single galaxy and full 2-dimensional (2D) coverage of the
disks.

In this paper, we describe the PPAK IFS Nearby Galaxies Survey: PINGS, a
project designed to construct 2D spectroscopic mosaics of a
representative sample of nearby spiral galaxies, using the unique instrumental
capabilities of the Postdam Multi Aperture Spectrograph, PMAS
\citep{Roth:2005p2463} in the PPAK mode
\citep{Verheijen:2004p2481,Kelz:2006p3341,Kelz:2006p338} at the Centro
Astron{\'o}mico Hispano Alem{\'a}n (CAHA) at Calar Alto, Spain. ``The PMAS fibre
PAcK (PPAK) is currently one of the world's widest integral field units with a
field-of-view (FOV) of 74\,$\times$\,65 arcsec that provides a semi-contiguous
regular sampling of extended astronomical objects'' (Kelz,
\url{http://tinyurl.com/ppak-aip}).
This project represents one of the first attempts to obtain 2D spectra of the whole
surface of a galaxy in the nearby universe. The observations consist of
integral field unit (IFU) spectroscopic mosaics for 17 nearby galaxies
($D\,<\,100$ Mpc) with a projected optical angular size of less than 10
arcmin. The spectroscopic mosaicking comprises more than 50000 spectra in the optical
wavelength range. The data set will be supplemented with broad band and narrow
band imaging for those objects without publicly available images.

The primary scientific objectives of PINGS are to use the 2D IFS observations
to study the small and intermediate scale variation in the line emission and
stellar continuum by means of pixel-resolved maps across the disks of nearby
galaxies. These spectral maps will allow us to test, confirm, and extend the
previous body of results from small-sample studies, while at the same time
open up a new frontier of studying the two-dimensional metallicity structure
of disks and the intrinsic dispersion in metallicity.
Furthermore, the large body of data arising from these studies will also allow
us to test and strengthen the diagnostic methods that are used to measure
\hh region abundances in galaxies.

Previous works have used multi-object instruments to obtain simultaneous 
spectra of \hh regions in a disk galaxy
\citep[e.g.][]{Roy:1988p320,Kennicutt:1996p1603,Moustakas:2006p313}, 
or narrow-band imaging of specific fields to obtain information of star
forming regions and the ionized gas \citep[e.g.][]{Scowen:1996p2663}.
One important attempt is represented by the SAURON project
\citep{Bacon:2001p2659}, which is based on a panoramic lenslet array
spectrograph with a relatively large FOV of 33\,$\times$\,41
arcsec$^2$. SAURON was specifically designed to study the kinematics and stellar
populations of a sample of nearby elliptical and lenticular galaxies. The
application of SAURON to spiral galaxies was restricted to the study of spiral bulges.
A recent effort by \citet{Rosolowsky:2008p1636} plans to obtain spectroscopy
for $\sim$\,1000 \hh regions through the M\,33 Metallicity Project, using
multi-slit observations. On the other hand, \citet{Blanc:2009p3483} obtained
IFS observations of the central region of M\,51 ($\sim$ 1.7 arcmin$^2$), using
the VIRUS-P instrument.
However, in spite of the obvious advantages of the IFS technique in tackling known
scientific problems and in opening up new lines of research, 2D spectroscopy
is a method that has been used relatively infrequently to study large
angular-size nearby objects.
Recently, PPAK was used successfully to map the Orion nebula, obtaining the
chemical composition through strong line ratios \citep{Sanchez:2007p1696}. 
Likewise, PMAS in the lens-array configuration was used to map the spatial
distribution of the physical properties of the dwarf \hh galaxy II\,Zw\,70
\citep{Kehrig:2008p1124}, although covering just a small FOV
($\sim$\,32 arcsec).

Despite these previous efforts toward IFS of nearby galaxies, the application
to obtain complete 2D information in galaxies is a novel technique. Reasons
for the lack of studies in this area include small wavelength coverage,
fibre-optic calibration problems, but mainly the limited FOV of the instruments
available worldwide. Most of these IFUs have a FOV of the order of arcsec,
preventing a good coverage of the target galaxies on the sky in a reasonable
time, even with a mosaicking technique. Furthermore, in some cases the
spectral coverage is not appropriate to measure important diagnostic
emission-lines used in chemical abundance studies. Moreover, the complex data
reduction and visualisation imposes a further obstacle to more ambitious
projects based on 2D spectroscopy. To our knowledge there has been no attempt
to obtain point-by-point spectra over a large wavelength range of the whole
surface of a galaxy covering all \hh regions within it. 
Similarly to SAURON for early-type galaxies, PINGS will provide the most
detailed knowledge of star formation and gas chemistry across a late-type
galaxy. This information is also relevant for interpreting the integrated
colours and spectra of high redshift sources. In that respect, PINGS
represents a leap in the study of the chemical abundances and the global
properties of galaxies.

The objectives of this paper are: 1) to provide the background information of the
PINGS survey; including a detailed description of the observations, 
all the data reduction techniques implemented (some of them novel in the
treatment of IFS data), and all the uncertainties involved in this process; 2) to
offer a general description of the procedures
involved in IFS observations of this kind; and 3) to present the PINGS data
products and the archival value of this survey.
The paper is organised as follows. In \S\,2 we describe the core scientific
objectives of the PINGS project and we discuss some of the many applications
of the data set. In \S\,3 we describe the properties of the galaxy sample
selected, while in \S\,4 we explain the observational strategy and the PINGS
observations themselves. In \S\,5 we present the des\-crip\-tion of the complex
data reduction involved in this IFS survey, while in \S\,6 we describe the
sources and magnitudes of the errors and uncertainties in the data sample.
In \S\,7, we summarise the basic properties of the data, showing
a few science case examples extracted from the data set, including the
integrated properties, emission line maps and a comparison of line intensity
ratios with previously published studies for some galaxies. Finally, in \S\,8
we give a summary of the article.

\section{Scientific Objectives}
\label{sec:science}

The study of chemical abundances in galaxies has been significantly benefited
from the vast amount of data collected in recent years, either at the
neighborhood of the Sun, or at high redshifts, especially on large scale
surveys such as the SDSS or the 2dF galaxy redshift survey
\citep{Colless:2001p2675}. Most studies derived from these observations
have focused on linking the properties of high redshift galaxies with nearby
objects, as an attempt to understand the principles of the formation and
chemical evolution of the galaxies \citep[e.g. see][]{Pettini:2006p3505}.
Historically, the metal content of low redshift galaxies has been determined
through the nebular emission of individual \hh regions at discrete spatial
positions, these measurements provide hints on the chemical evolution,
stellar nucleosynthesis and star formation histories of spiral galaxies.
The chemical evolution is dictated by a complex
array of parameters, including  the local initial composition, the
distribution of molecular and neutral gas, star formation
history (SFH), gas infall and outflows, radial transport and mixing of gas
within disks, stellar yields, and the initial mass function (IMF). Although it
is difficult to disentangle the effects of the various contributors, measurements
of current elemental abundances constrain the possible evolutionary histories
of the existing stars and galaxies, and thus the importance of the accurate
determination of the chemical composition among different galaxy types.

Different studies have shown a complex link between the chemical abundances of
galaxies and their physical properties. Such studies are only able to
accurately measure the first two moments of the abundance distribution --the
mean metal abundances of disks and their radial gradients-- and on
characterising the relations between these abundance properties and the
physical properties of the parent ga\-la\-xies, for example galactic luminosity,
stellar and dynamical mass, circular velocity, surface brightness, colors,
mass-to-light ratios, Hubble type, gas fraction of the disk, etc.
These studies have revealed a number of important scaling laws and systematic
patterns including luminosity-metallicity, mass-metallicity, and surface
brightness vs. metallicity relations
\citep[e.g.][]{Skillman:1989p1592,VilaCostas:1992p322,Zaritsky:1994p333,Tremonti:2004p1138},
effective yield vs. luminosity and circular velocity relations
\citep[e.g.][]{Garnett:2002p339}, abundance gradients and the effective radius
of disks \citep[e.g.][]{Diaz:1989p3307}, and systematic differences in the gas-phase
abundance gradients between normal and barred spirals
\citep[e.g.][]{Zaritsky:1994p333,Martin:1994p1602}.
However, these studies have been limited by the number of objects sampled, the
number of \hh regions observed and the coverage of these regions within the
galaxy surface.

In order to obtain a deeper insight of the mechanisms that rule the chemical
evolution of galaxies, we require the combination of high quality 
multi-wavelength data and wide field optical spectroscopy in order to increase
significantly the number of \hh regions sampled in any given galaxy. 
The PINGS project was conceived to tackle the problem of the 2-dimensional
coverage of the whole galaxy surface. The imaging spectroscopy technique
applied in PINGS provides a powerful tool for studying the distribution of
physical properties in nearby well-resolved galaxies.
PINGS was specially designed to obtain complete maps of the
emission-line abundances, stellar populations, and reddening using an IFS
mosaicking {\em imaging}, which takes advantage of what is currently one of the
world's widest FOV IFU.
With this novel spectroscopic technique, the data can be used to derive: 1)
oxygen abundance distributions based on a suite
of strong-line diagnostics incorporating absorption-corrected H$\alpha$, H$\beta$,
[O\,{\footnotesize II}], [O\,{\footnotesize III}], [N\,{\footnotesize II}],
and [S\,{\footnotesize II}] line ratios; 2) local nebular
reddening estimates based on the Balmer decrement; 3) measurements of
ionization structure in \hh regions and diffuse ionized gas using the
well-known and most updated forbidden-line diagnostics in the oxygen 
and nitrogen lines; 4) rough fits to the stellar age mix from the stellar spectra.

The resulting spectral maps and ancillary data will be used to address
a number of important astrophysical issues regarding both the gas-phase and
the stellar populations in galaxies. For example, one application will be able
to test whether the metal abundance distributions in disks are
axisymmetric. This is usually taken for granted in chemical evolution models,
but one might expect strong deviations from symmetry in strongly lopsided,
interacting, or barred galaxies, which are subject to large scale gas
flows. Another important goal is to place strong limits on the dispersion in
metal abundance locally in disks; there is evidence for a large dispersion in
some objects such as NGC\,925 or M\,33 \citep{Rosolowsky:2008p1636}, but it is
not clear from those data whether the dispersion is due to non-axisymmetric
abundance variations, systematic errors in the abundance measurements, or a
real local dispersion. 
Yet another by product of our analysis will be point by point reddening maps of
the galaxies, which can be combined with UV, H$\alpha$, and infrared maps to
derive robust, extinction-corrected maps of the SFR.

PINGS can also provide a very detailed knowledge of the role played by star
formation in the cosmic life of galaxies. All the important scaling laws
previously mentioned tell us that, once born, stars change the ionization state, the
kinematics and chemistry of the interstellar medium and, thus, change the
initial conditions of the next episode of star formation.
Substantially, star formation is a loop mechanism which drives the luminosity,
mass and chemical evolution of each galaxy (leaving aside external agents like
interactions and mergers). The details of such a complex mechanism are still
not well established observationally and not well developed theoretically,
and limit our understanding of galaxy evolution from the early universe to
present day.
In combination with ancillary data, the flux maps computed from the PINGS data
will be used to study both the most recent star formation activity of the
targets and the older stellar populations. We will be able to
identify the gas and stellar features responsible for the observed spectra, to
derive the dependence of the local star formation rate on the local surface
brightness, a key recipe for modelling galaxy evolution and the environmental
dependence of star formation. These data will also provide an important check for
interpreting the integrated broad-band colours and spectra of high redshift
sources.

\begin{figure*}
  \includegraphics[width=0.9\textwidth]{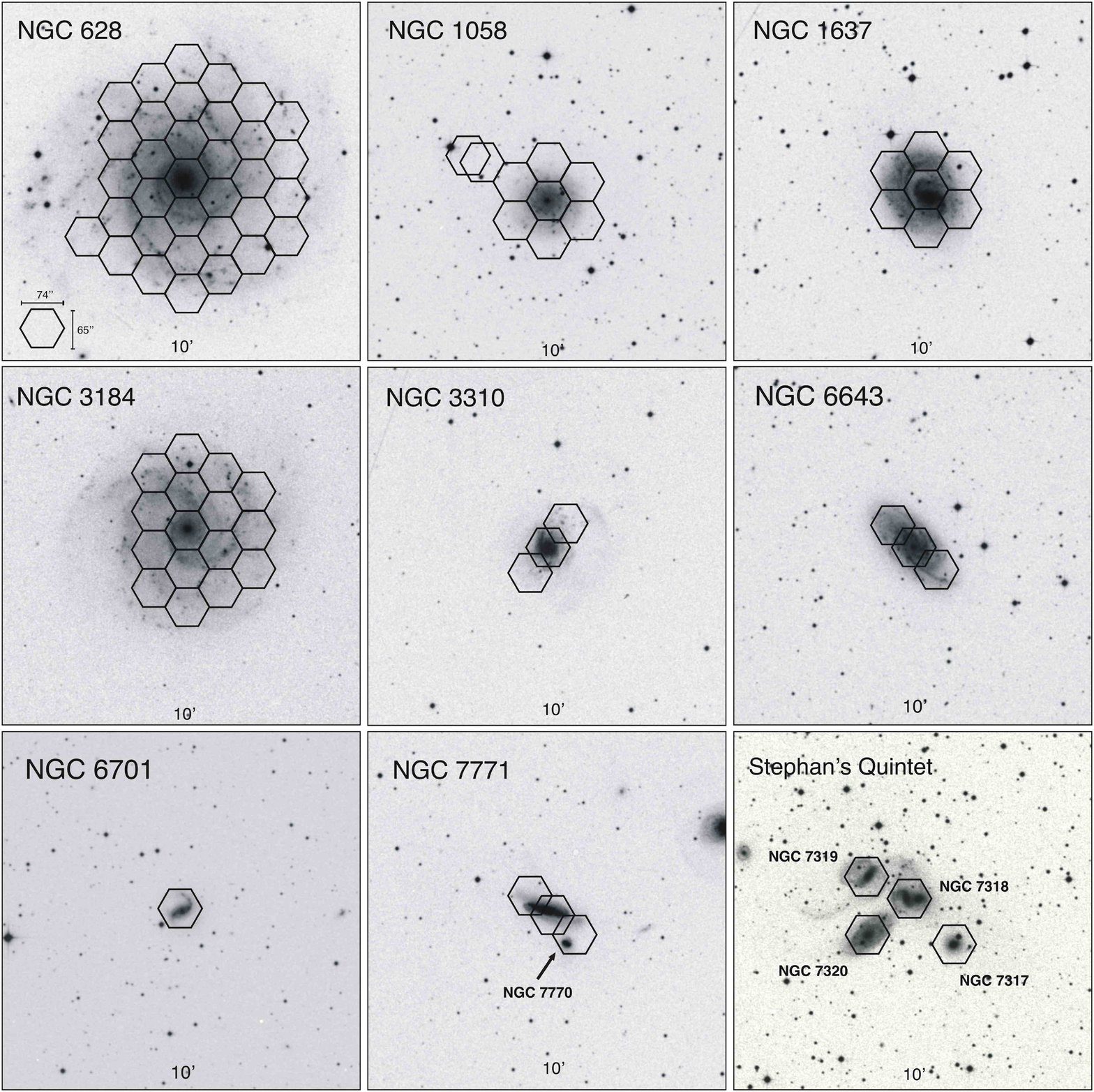}
  \caption{Digital Sky Survey images for a selection of objects included in the PINGS
    sample. On each image, an array of hexagonal fields corresponding to the
    PPAK aperture is superimposed on scale, showing the IFU mosaicking
    technique and the observed positions. All images are 10'\,$\times$\,10' and
    displayed in top-north, left-east configuration.}
  \label{fig:sample}
\end{figure*}

\section{Sample description}
\label{sec:sample}

In order to achieve the scientific goals described above, we incorporated a
diverse population of galaxies, adopting a physically based
approach to defining the PINGS sample. We decided to observe a set of local
spiral galaxies which were representative of different galaxy types.
However, the size and precise nature of the sample were heavily influenced by
a set of technical considerations, the principal limiting factor being the
FOV of the PPAK unit. We wanted to observe relatively nearby
galaxies to maximise the physical linear resolution using the mosaicking
technique. However, we also had to take into account the limitations imposed by the
amount of non-secure observing time and meteorological conditions for the
granted runs. Therefore, the sample size was dictated by a balance between
achieving a representative range of galaxies properties and practical
limitations in observing time.

When constructing the sample, we also took into account a range
of other galaxy properties, such as inclination (with preference to face-on
galaxies), surface brightness, bar structure, spiral arm structure, and
environment (i.e. isolated, interacting and clustered). We favoured galaxies
with high surface brightness and active star formation so that we could have a
good distribution of \hh regions across the galaxies surface. Galaxies with
bars and/or non-typical spiral morphology were also preferred.
The final selection of
galaxies also took into account practical factors such as optimal equatorial
right ascension and declination for the location of Calar Alto observatory,
and the ob\-ser\-va\-ble time per night for a given object above a certain
airmass (due to problems of differential atmospheric refraction).

The PINGS sample consists of 17 galaxies within a ma\-xi\-mum distance of 100
Mpc; the average distance of the sample is 28 Mpc (for $H_0$ = 73
km\,s$^{-1}$\,Mpc$^{-1}$). The final sample includes normal, lopsided,
interacting and barred spirals with a good range of galactic properties and SF
environments with multi-wavelength pu\-blic data. 
A good fraction of the sample belongs to the
Spitzer Infrared Nearby Galaxies Survey \citep[SINGS,][]{Kennicutt:2003p1560}, 
which ensures a rich set of ancillary observations in the UV, infrared, \hi
and radio.

The sample objects were given a different observing priority based on the
angular size of the objects, the number of PPAK adjacent pointings necessary
to complete the mosaic, and the scientific relevance of the galaxy. 
\autoref{tab:sample} gives a complete listing of the PINGS sample with some
of their relevant properties. The first priority
was assigned to medium-size targets such as NGC\,1058, NGC\,1637, NGC\,3310,
NGC\,4625 and NGC\,5474 which are bright, face-on spirals of very different
morphological type, with many sources of ancillary data and could be covered
with relatively few IFU pointings. The second prio\-ri\-ty was
given to smaller galaxies which fit perfectly in terms of size and acquisition time for the
periods during the night when the first priority objects were not observable (due to
a high airmass or bad weather conditions) and/or in the case their mosaicking
was completed.

NGC\,628 (Messier\,74) is a special object among the selected
galaxies and the most interesting one of the sample. NGC\,628 is a close,
bright, grand-design spiral galaxy which has been extensively studied. 
With a projected optical size of 10.5\,$\times$\,9.5 arcmin, it is the most
extended object of the sample. Although it could be considered too large
to be fully observed in a realistic time, we attempted the observation of this
galaxy considering that the spectroscopic mosaicking of NGC\,628 represents
the real 2-dimensional scientific spirit of the PINGS project (see S\'anchez
et al. 2009, hereafter Paper II).
Such a large galaxy would offer us the possibility to assess the
body of results from the rest of the small galaxies in the sample and would
allow us to study the 2D metallicity structure of the disk and second
order properties of the abundance distribution.
The observations of this galaxy spanned a period of three-years. Hitherto,
NGC\,628 represents the largest area ever covered by an IFU, as described
briefly in the next section and in detail in \citetalias{paperII}.
A special priority was also given to NGC\,3184, galaxy which falls between the
medium first-priority and large size galaxies. The observations for this
object spanned for 2 years, obtaining an almost complete mosaicking (see
\autoref{fig:sample}), making this galaxy the 2nd largest object of the
sample.

\autoref{fig:sample} shows Digital Sky Survey\footnote{The Digitized Sky
  Survey was produced at the Space Telescope Science
  Institute under U.S. Government grant NAG W-2166. The images of these
  surveys are based on photographic data obtained using the Oschin Schmidt
  Telescope on Palomar Mountain and the UK Schmidt Telescope. The plates
  were processed into the present compressed digital form with the
  permission of these institutions.} images for a selection of galaxies
listed in \autoref{tab:sample}. The mosaicking of the largest objects in the
sample, NGC\,628 and NGC\,3184, consist of 34 and 16 individual IFU pointings
respectively, covering almost completely the spiral arms of these two bright
grand-design galaxies. The outlying pointings of NGC\,1058 and the mosaicking
configurations of NGC\,3310, NGC\,6643 and NGC\,7771 are explained in the next
section.

In summary, the PINGS sample was selected in a careful way to find
a trade-off between the size of the galaxies, their morphological types, their
physical properties and the practical limitations imposed by the instrument
and the amount of observing time. The result is a comprehensive sample of
galaxies with a good range of galactic properties and available multi-wavelength
ancillary data, in order to maximise both the original science goals of the
project and the possible archival value of the survey.

\begin{table*}
\label{tab:sample}
\begin{minipage}{0.9\textwidth}
\caption[Galaxy properties]{Galaxy properties of the PINGS sample. 
  Col.\,(1): Galaxy name. 
  Col.\,(2):
  Morphological type from the R3C catalog\citep{deVaucouleurs:1991p2393}. 
  Col.\,(3): 
  Distances in Mpc, references:
  NGC\,628, \citet{Hendry:2005p2408}; 
  NGC\,1058, \citet{Eastman:1996p2401};
  NGC\,1637, \citet{Saha:2006p2402}; 
  NGC\,2976, \citet{Karachentsev:2002p3342};
  NGC\,3184, \citet{Leonard:2002p2403};
  NGC\,3310, \citet{Terry:2002p2434}; 
  NGC\,5474, \citet{Drozdovsky:2000p2406};
  NGC\,6643, \citet{Willick:1997p2431}; 
  NGC\,4625, NGC\,6701, NGC\,7771, \& Stephan's Quintet: Galactocentric GSR
  distances derived from the redshift, assumming a value of $H_0$ = 73
  km\,s$^{-1}$ Mpc$^{-1}$.
  Col.\,(4): 
  Projected size, major and minor axes at the $B_{25}$ mag
  arcsec$^{-2}$ from R3C, except NGC\,7318 from \citet{Jarrett:2003p2447}. 
  Col.\,(5): 
  Absolute $B$-band magnitude calculated from the apparent
  magnitude listed in the R3C catalog and the adopted distances to the
  system. 
  Col.\,(6): 
  Redshift, references: 
  NGC\,628, \citet{Lu:1993p2436}; 
  NGC\,4625, \citet{Fisher:1995p2440}; 
  NGC\,6701, \citet{Theureau:1998p2442};
  NGC\,7770, \citet{Woods:2006p2443};
  NGC\,1637, NGC\,3310, \citet{Haynes:1998p2438}; 
  NGC\,2976, NGC\,5474, \citet[The Updated Zwicky Catalog]{Falco:1999p3347};
  NGC\,1058, NGC\,3184, NGC\,6643, NGC\,7771, \citet{Springob:2005p3356};
  NGC\,7317, NGC\,7318a, NGC\,7318b, \citet{Hickson:1992p2445}; 
  NGC\,7319, NGC\,7320, \citet{Nishiura:2000p2446}. 
  Col.\,(7): 
  Heliocentric velocities calculated from $v=zc$, with no further correction
  applied.
  Col.\,(8): 
  Galaxy inclination angle based on the $B_{25}$ mag arcsec$^{-2}$ from R3C. 
  Col.\,(9): 
  Galaxy position angle, measured positive NE, in the $B_{25}$
  mag arcsec$^{-2}$ except for NGC\,1058, NGC\,1637, NGC\,7317, NGC\,7319,
  NGC\,7770, which are based on the $K_s$-band \citep{Jarrett:2003p2447}, and
  NGC\,3310, NGC\,4625, NGC\,5474 based on the $r$-SDSS band. 
  Col.\,(10): Galaxy location.
}

\begin{tabular}{@{\extracolsep{\fill}} lccc D{.}{.}{3.2} D{.}{.}{1.4} D{.}{}{4.0} ccc }

\hline

& & \multicolumn{1}{c}{Distance} &  Projected size  &&& \multicolumn{1}{c}{$V_{\odot}$}  \\
\multicolumn{1}{c}{Object} & Type & \multicolumn{1}{c}{{\scriptsize (Mpc)}} & {\scriptsize (arcmin)} & 
\multicolumn{1}{c}{$M_B$} & \multicolumn{1}{c}{$z$} & \multicolumn{1}{c}{{\scriptsize (km\,s$^{-1}$)}} & $i$ & P.A. & Constellation \\

\multicolumn{1}{c}{{\scriptsize (1)}} &
\multicolumn{1}{c}{{\scriptsize (2)}} &
\multicolumn{1}{c}{{\scriptsize (3)}} &
\multicolumn{1}{c}{{\scriptsize (4)}} &
\multicolumn{1}{c}{{\scriptsize (5)}} &
\multicolumn{1}{c}{{\scriptsize (6)}} &
\multicolumn{1}{c}{{\scriptsize (7)}} &
\multicolumn{1}{c}{{\scriptsize (8)}} &
\multicolumn{1}{c}{{\scriptsize (9)}} &
\multicolumn{1}{c}{{\scriptsize (10)}} \\

\hline

 NGC\,628\,\dotfill  & SA(s)c       &  9.3 & 10.5\,$\times$\,9.5 & -19.9 & 0.00219 &  657. & 24 &   25 & Pisces \\ [2pt]

 NGC\,1058\,\dotfill & SA(rs)c      & 10.6 &  3.0\,$\times$\,2.8 & -18.3 & 0.00173 &  519. & 21 &   95 & Perseus \\ [2pt]

 NGC\,1637\,\dotfill & SAB(rs)c     & 12.0 &  4.0\,$\times$\,3.2 & -18.9 & 0.00239 &  717. & 36 &   33 & Eridanus \\ [2pt]

 NGC\,2976\,\dotfill & SAc pec      &  3.6 &  5.9\,$\times$\,2.7 & -16.9 & 0.00008 &   24. & 63 &  143 & Ursa Major \\ [2pt]

 NGC\,3184\,\dotfill & SAB(rs)cd    & 11.1 &  7.4\,$\times$\,6.9 & -19.9 & 0.00194 &  582. & 21 &  135 & Ursa Major \\ [2pt]

 NGC\,3310\,\dotfill & SAB(r)bc     & 17.5 &  3.1\,$\times$\,2.4 & -20.1 & 0.00331 &  993. & 39 &  163 & Ursa Major \\ [2pt]

 NGC\,4625\,\dotfill & SAB(rs)m     &  9.0 &  2.2\,$\times$\,1.9 & -16.9 & 0.00203 &  609. & 29 &   30 & C.~Venatici \\ [2pt]

 NGC\,5474\,\dotfill & SA(s)cd      &  6.8 &  4.8\,$\times$\,4.3 & -17.9 & 0.00098 &  294. & 27 &   91 & Ursa Major \\ [2pt]

 NGC\,6643\,\dotfill & SA(rs)c      & 20.1 &  3.8\,$\times$\,1.9 & -19.8 & 0.00495 & 1485. & 60 &   37 & Draco \\ [2pt]

 NGC\,6701\,\dotfill & SB(s)a       & 57.2 &  1.5\,$\times$\,1.3 & -20.8 & 0.01323 & 3969. & 32 &   24 & Draco  \\ [2pt]

 NGC\,7770\dotfill   & S0           & 58.7 &  0.8\,$\times$\,0.7 & -19.4 & 0.01414 & 4242. & 27 &   50 & Pegasus   \\ [2pt]

 NGC\,7771\,\dotfill & SB(s)a       & 60.8 &  2.5\,$\times$\,1.0 & -20.8 & 0.01445 & 4335. & 66 &   68 & \multicolumn{1}{c}{"} \\ [2pt]

 Stephan's Quintet  &&&&&&&&& Pegasus \\

\hspace{5pt} NGC\,7317\,\dotfill  & E4          & 93.3 & 1.1\,$\times$\,1.1 & -20.3 &  0.02201 &  6603. & 12 & 150 &\multicolumn{1}{c}{"} \\ [2pt]

\hspace{5pt} NGC\,7318A\,\dotfill & E2 pec      & 93.7 & 0.9\,$\times$\,0.9 & -20.5 &  0.02211 &  6633. & \ldots & \ldots &\multicolumn{1}{c}{"} \\ [2pt]

\hspace{5pt} NGC\,7318B\,\dotfill & SB(s)bc pec & 82.0 & 1.9\,$\times$\,1.2 & -20.6 &  0.01926 &  5778. & \ldots & \ldots &\multicolumn{1}{c}{"} \\ [2pt]

\hspace{5pt} NGC\,7319\,\dotfill  & SB(s)bc pec & 95.4 & 1.7\,$\times$\,1.3 & -20.8 &  0.02251 &  6753. & 41 & 148 &\multicolumn{1}{c}{"} \\ [2pt]

\hspace{5pt} NGC\,7320\,\dotfill  & SA(s)d      & 13.7 & 2.2\,$\times$\,1.1 & -17.5 &  0.00262 &   786. & 59 & 132 &\multicolumn{1}{c}{"} \\ [2pt]

\hline
\end{tabular}
\end{minipage}
\end{table*}

%
%
%

%
%
%

\section{Observations}
\label{sec:obs}

Observations for the PINGS galaxies were carried out at the 3.5m telescope of
the Calar Alto observatory with the Postdam Multi Aperture Spectrograph, PMAS
\citep{Roth:2005p2463} in the PPAK mode
\citep{Verheijen:2004p2481,Kelz:2006p3341}, i.e. ``a retrofitted bare fibre
bundle IFU which expands the FOV of PMAS to a hexagonal area with a footprint
of 74\,$\times$\,65 arcsec, with a filling factor of 65\% due to gaps in
between the fibres''.
The PPAK unit features a central hexagonal bundle with 331 densely packed optical
fibres to sample an astronomical object at 2.7 arcsec per fibre. 
The sky background is sampled by 36 additional fibres distributed in 6
mini-IFU bundles of 6 fibres each, in a circular
distribution at $\sim$\,90 arcsec of the centre and at the edges of the central
hexagon. The sky-fibres are distributed among the science ones in the pseudo-slit, in
order to have a good characterisation of the sky. Additionally, 15 fibres can
be illuminated directly by internal lamps to calibrate the instrument.

All sample galaxies were observed using the same telescope and instrument
set-up. We used the V300 grating, covering a wavelength between 3700\,--\,7100
\AA\ with a resolution of $\sim$\,10 \AA\ FWHM, corresponding to $\sim$\,600
km\,s$^{-1}$. With this set-up we cover all the optical strong emission lines
used in typical abundance diagnostic methods \citep{Sanchez:2007p1696}.
For the particular setup used in the PINGS survey there was no need to 
use a order separating filter. The main reason being that the efficiency of
the instrument + telescope system (mostly the grating reflectivity and the
fibre transmission), drops dramatically at wavelengths shorter than $\sim$
3600\,\AA, where the transmission is 1/40000 of the one at the peak intensity
for the V300 grating ($\sim$ 5400 \AA) and 1/10000 of the value at redder
wavelengths covered by our survey ($\sim$ 6500\,\AA). At bluer wavelengths
the efficiency is even lower. Therefore, the system by itself blocks any 
possible 2nd order contamination up to $\sim$ 7200 \AA, and only at larger 
wavelengths it is required an order separation filter \citep{Kelz:2006p3341}.
The exposure times were calculated from previous experience with the
instrument in order to obtain spectroscopy with S/N $\ge$ 20 in the con\-ti\-nuum
and S/N $\ge$ 50 in the H$\alpha$ emission line for the brightest \hh
regions with the given grating.

Different observing strategies were implemented depending on the size and
priority of the targets. For those objects with relatively small angular size,
single PPAK pointings would not sample the surface of the galaxy with enough
spatial resolution, due to the incomplete filling factor of the fibre
bundle. In this case, a dithering procedure was applied
\citep{Sanchez:2007p3299}. For each individual position in dithering mode, the
first exposure was recorded and then, two consecutive exposures with the same
acquisition time were recorded, but with small offsets of
$\Delta$(RA,\,Dec)\,=\,(1.56,\,0.78) and (1.56,\,--0.78) arcsec with respect
to the first exposure. The advantage of this method is that all gaps of the
original exposure are covered, and every single point of the dithered field is
spectroscopically sampled within the resolution. The pitfalls are that the
exposure time and the amount of data to be processed is triple a normal
frame, preventing the possibility of applying this method to large mosaics.
We used a mean acquisition time per PPAK field in dithering mode (including
set-up + integration time) of 2\,$\times$\,600 sec. per dithering position for a
total of 60 min. exposure; and 3\,$\times$\,600 sec. for non-dithered frames.

The observations extended over a period of three-years with a total of 19
observing nights distributed during different observing runs and seasons.
For all the objects in the sample, the first exposure was centered in a
given geometrical position which, depending on the morphology or a
particular mosaicking pattern, may or may not coincide with the bright bulge of
the galaxy.
Consecutive pointings followed in ge\-ne\-ral a hexagonal pattern, adjusting
the mosaic pointings to the shape of the PPAK science bundle as shown in
\autoref{fig:sample}. Due to the shape of the PPAK bundle and by
construction of the mosaics, 11 spectra of each pointing corresponding to one
edge of the hexagon, overlap with the same number of spectra from the previous
pointing. This pattern was selected to maximise the covered area, but to allow
enough overlapping to match the different exposures taken
under variable atmospheric conditions. Exceptions are NGC\,2976, NGC\,3310,
NGC\,6643 and NGC\,7770 in which the mosaics were constructed to optimise the
galaxy surface coverage as explained below.

\begin{figure*}
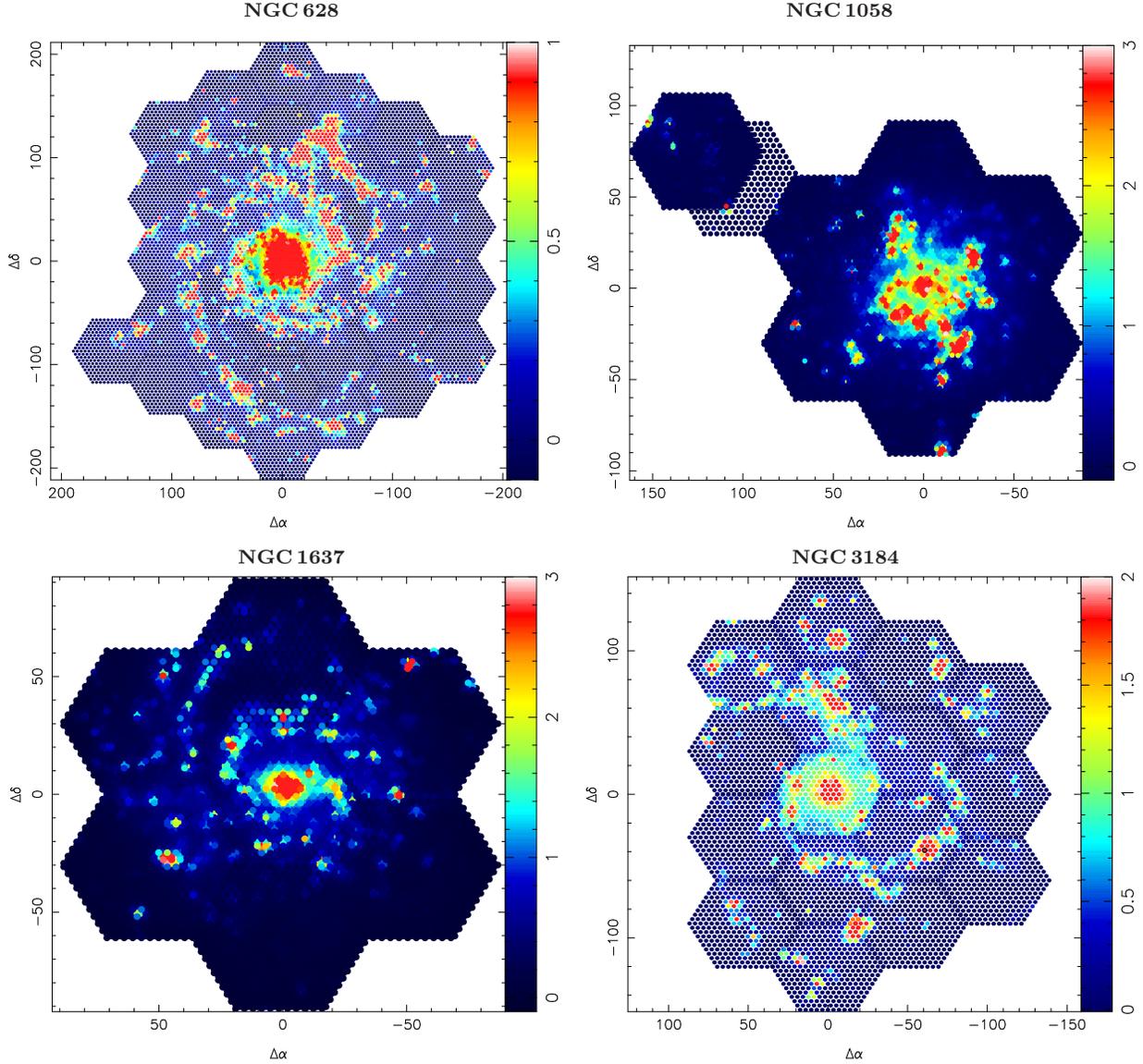

  \begin{minipage}{\textwidth}
    \centering
    \bf NGC\,628 \hspace{0.35\textwidth} NGC\,1058
    \vspace{-0.3cm}
  \end{minipage}
  \includegraphics[width=7cm,angle=-90]{n0628_map}\hspace{0.3cm}
  \includegraphics[width=7cm,angle=-90]{n1058_map}
  \begin{minipage}{\textwidth}
  \vspace{0.3cm}
  \centering
    \bf NGC\,1637 \hspace{0.35\textwidth} NGC\,3184
  \vspace{-0.1cm}
  \end{minipage}
  \vspace{0.3cm}
  \includegraphics[width=7cm,angle=-90]{n1637_map}\hspace{0.3cm}
  \includegraphics[width=7cm,angle=-90]{n3184_map}
  \caption{Examples of the spectroscopic mosaics of NGC\,628, NGC\,1058, NGC\,1637
    and NGC\,3184. Each panel shows an intensity level {\it narrow-band} map
    centered at H$\alpha$ (6563 \AA) in units of 10$^{-16}$
    erg\,s$^{-1}$\,cm$^{-2}$\,arcsec$^{-2}$. Note the effect of surface area
    coverage for dithered (e.g. NGC\,1058, NGC\,1637) vs. non-dithered
    (e.g. NGC\,628, NGC\,3184) observations, where the gaps between the fibres
    are clearly seen.}
  \label{fig:maps_1}
\end{figure*}

\begin{figure*}
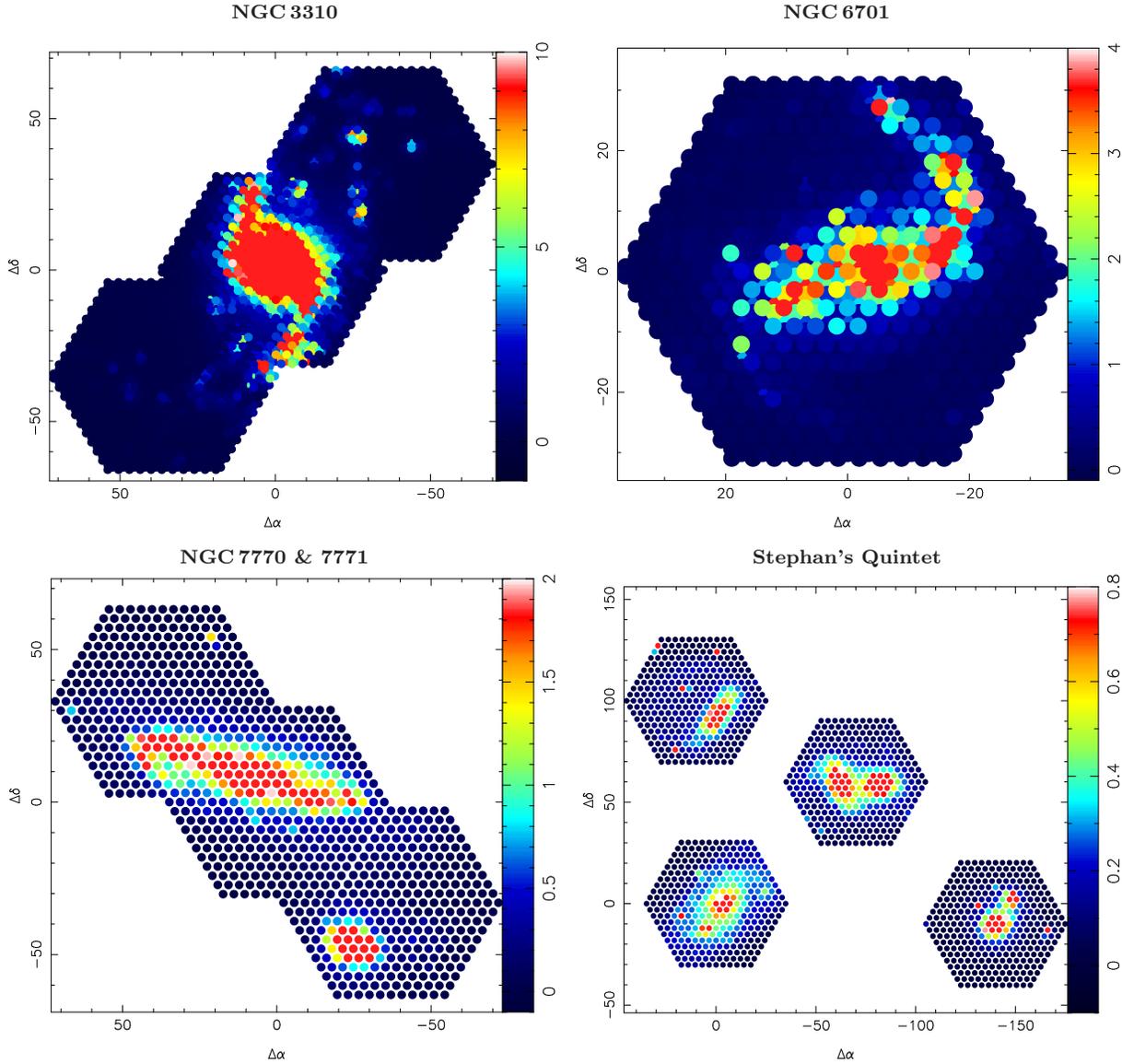

  \begin{minipage}{\textwidth}
    \centering
    \bf NGC\,3310 \hspace{0.35\textwidth} NGC\,6701
    \vspace{-0.3cm}
  \end{minipage}
  \includegraphics[width=7cm,angle=-90]{n3310_map}\hspace{0.3cm}
  \includegraphics[width=7cm,angle=-90]{n6701_map}
  \begin{minipage}{\textwidth}
  \vspace{0.3cm}
  \centering
    \bf NGC\,7770 \& 7771 \hspace{0.3\textwidth} Stephan's Quintet
  \vspace{-0.1cm}
  \end{minipage}
  \vspace{0.3cm}
  \includegraphics[width=7cm,angle=-90]{n7771_map}\hspace{0.3cm}
  \includegraphics[width=7cm,angle=-90]{stephan_map}
  \caption{Spectroscopic mosaics of NGC\,3310, NGC\,6701, NGC\,7770, and the Stephan's
    Quintet. Note the effect of dithered (top panels) vs. non-dithered (bottom panels)
    observations. Units as in \autoref{fig:maps_1}.}
  \label{fig:maps_2}
\end{figure*}

\subsection{NGC 628}

NGC\,628 (or M\,74) is an extensively studied isolated grand-design
Sc spiral galaxy at a distance of 9.3 Mpc in the constellation of Pisces. The
observations for this galaxy totaled six observing nights and 34 different
pointings. The central position was observed in dithering
mode to gain spatial resolution, while the remaining 33 positions were
observed without dithering due to the large size of the mosaic. Seven
positions were observed on the 28th October 2006, 19 positions were
observed between the 10th and 12th of December 2007, 1 position on August 9th 2008
and the remaining pointings on October 28th 2008. \autoref{fig:sample} shows
the mosaic pattern covering NGC\,628 consisting in a central position and
consecutive hexagonal concentric rings. The area covered by all the observed
positions accounts approximately for 34 arcmin$^2$, making this galaxy the
largest area ever covered by a IFU mosaicking. The spectroscopic mosaic
contains 11094 individual spectra, considering overlapping and repeated
exposures \citepalias[see][]{paperII}.

\subsection{NGC 1058}

NGC\,1058 is a well studied Sc spiral with a projected size of
3.0\,$\times$\,2.8 arcmin at a distance of 10.6 Mpc, in the constellation of Perseus.
The observations for this galaxy were performed on three consecutive nights
from the 7th to the 9th December 2007. The mosaic consists of the central position and one
concentric ring, covering most of the galaxy surface within one optical radius
(defined by the $B$-band 25th magnitude isophote). \citet{Ferguson:1998p224}
found the existence of \hh regions out to and beyond two optical radii in this
galaxy. We tried to observe these intrinsically interesting objects by performing a
couple of offsets of 2 and 2.5 arcmin north-east from the central position. These
2 additional position were merged to the original 7 tiles for a mosaic,
covering an area of approximately 8.5 arcmin$^2$ (see
\autoref{fig:sample}). All positions (with the exception of one blind offset)
were observed in dithering mode, accounting for a spectroscopic mosaic
containing 7944 individual spectra. At the time of the observations, we were
able to observe the recently discovered supernova 2007gr, a SN type Ic
located at 24''.8 west and 15''.8 north of the nucleus of NGC 1058 between two
foreground stars (see \autoref{sec:sn} and \autoref{fig:integ}).

\subsection{NGC 1637}

NGC\,1637 is a SAB distorted galaxy in Eridanus with a projected size of
4.0\,$\times$\,3.2 arcmin, at a distance of 12 Mpc. This galaxy presents a
clear asymmetry with a third well-defined arm seen in optical images, an
unusually extended \hi envelope ($D_{\hi}/D_{25}$ = 3.0), and an optical centre
that differs from the kinematic centre by 9 arcsec \citep{Roberts:2001p3212}.
NGC\,1637 was observed during December 8th to 10th 2007. The mosaic was built
with a central position and one concentric ring of 6 pointings (see
\autoref{fig:sample}), covering most of the galaxy surface within one optical
radius. The mosaic covers approximately 7 arcmin$^2$. This galaxy has a full
spectroscopic mosaic containing 6951 individual spectra.

\begin{table}
\label{tab:observations}
\caption[Sumamry of observations of the PINGS sample]{Summary of observations of the PINGS
  sample. Col.\,(1): Galaxy name. Col.\,(2): Number of individual IFU positions
  observed for each galaxy mosaic. Col.\,(3): Status of the
  mosaicking; when a percentage is shown, the number in parentheses represents
  the total number of pointings necessary to cover the optical surface of the
  galaxy. Col.\,(4): Total number of spectra for each mosaic. Col.\,(5):
  Comments: a) Largest area ever covered by a IFU mosaicking, with a total of
  $\sim$\,35 arcmin$^2$; b) An offset position of NGC\,1058 was not observed in
  dithering mode (see \autoref{fig:maps_1}); c) Galaxy with the second largest
  area of the sample; d) NGC\,7770 within the field; e) NGC\,7318A and
  NGC\,7318B are included in one field; $^{\star}$All pointings observed in
  dithering mode.}

\begin{tabular}{@{} lcc  D{.}{}{5.0} c }
\hline
\multicolumn{1}{c}{Object} & Positions & Mosaic & \multicolumn{1}{c}{Spectra} & Notes \\

\multicolumn{1}{c}{{\scriptsize (1)}} &
\multicolumn{1}{c}{{\scriptsize (2)}} &
\multicolumn{1}{c}{{\scriptsize (3)}} &
\multicolumn{1}{c}{{\scriptsize (4)}} &
\multicolumn{1}{c}{{\scriptsize (5)}} \\

\hline

NGC\,628  &    34 & 92\% (37) & 13571 & $a$ \\ [2pt]

NGC\,1058 &     9 &  complete &  7944 & $\star$,$b$ \\ [2pt]

NGC\,1637 &     7 &  complete &  6951 & $\star$ \\ [2pt]

NGC\,2976 &     2 &  22\% (9) &   662 &  \\ [2pt]
 
NGC\,3184 &    16 & 84\% (19) &  5296 & $c$ \\ [2pt]

NGC\,3310 &     3 &  complete &  2979 & $\star$ \\ [2pt]

NGC\,4625 &     1 &  14\% (7) &   993 & $\star$ \\ [2pt]

NGC\,5474 &     6 &  86\% (7) &  5958 & $\star$ \\ [2pt]

NGC\,6643 &     3 & complete &  2979 & $\star$ \\ [2pt]

NGC\,6701 &     1 & complete &   993 & $\star$ \\ [2pt] 

NGC\,7771 &     3 & complete &   993 &  $d$ \\ [2pt]

Stephan's Q. &  4 & complete &  1324 &  $e$ \\ [2pt]

\hline
\end{tabular}
\end{table}

%
%
%

%
%
%

\subsection{NGC 2976}

NGC\,2976 is a SAc peculiar spiral galaxy with strong emission line spectra
with a projected size of 5.9\,$\times$\,2.7 arcmin in Ursa Major, at a
distance of 3.6 Mpc, being the closest object of the sample.
The observations for NGC\,2976 were carried out on October 30th 2008. Given
the distorted morphology of the galaxy a more convenient mosaic pattern was
designed. Two pointings were observed for this object, corresponding
to the central region of NGC\,2976. The observations were performed in
non-dithering mode. The spectroscopic data for this galaxy consist of 662
individual spectra.

\subsection{NGC 3184}

NGC\,3184, a SAB face-on galaxy located in Ursa Major, has the 2nd largest
angular size in the sample. It covers an area of 7.4\,$\times$\,6.9 arcmin at
a distance of 11 Mpc.
NGC\,3184 has been classified as one of the metal-richest galaxies ever observed
\citep{McCall:1985p1243,vanZee:1998p81}, and which has also harboured recently a
supernova explosion (SN 1999gi) \citep{Nakano:1999p3213}.
Three concentric rings are necessary to cover the entire optical
disk. Observations for this galaxy were performed on December
10th 2007, following the standard mosaicking pattern with a central position
and one complete ring of 7 IFU pointings. Then, on April 27th and 28th
2009, 9 additional pointings were observed covering partially a second
concentric ring as shown in \autoref{fig:sample}. The area covered by all the
observed positions is $\sim$\,16 arcmin$^2$. The spectroscopic data for this
galaxy consists of 5296 individual spectra.

\subsection{NGC 3310}

NGC\,3310 is a very distorted spiral galaxy with strong star formation in the
constellation of Ursa Major, at a distance of 17.5 Mpc. It covers an area of
3.1\,$\times$\,2.4 arcmin in the optical $B$-band, with a very bright central
nucleus, surrounded by a ring of luminous \hh regions.
Different studies of this galaxy suggest a recent merging episode which
triggered the burst of star formation \citep[][and references atherein]{Kregel:2001p3214,Wehner:2006p84}.
Given its morphology, a tailored mosaic pattern was
constructed for this galaxy (see \autoref{fig:sample}). Three pointings cover
the surface of NGC\,3310 with a central position centered in the galaxy's
nucleus and two offsets of (--35,\,35) and (35,\,--35) arcsec in (RA,\,Dec) in
north-west and south-east directions respectively. The observations were
carried out on December 8th 2007, and were performed in dithering mode. This
galaxy has a full spectroscopic mosaic, which covers an area of approximately
2.8 arcmin$^2$.  The spectroscopic data for this galaxy totals 2979
individual spectra.

\subsection{NGC 4625}

NGC\,4625 is a low-luminosity SAB, one-armed Magellanic spiral galaxy
thought to be interacting with the also single-armed spiral
NGC\,4618 in Canes Venatici, at a distance of 9 Mpc. The optical size of this
galaxy covers an area of approximately 2.2\,$\times$\,1.9 arcmin, however
Gil~de~Paz et al. (\citeyear{GildePaz:2005p2505})
discovered an extended UV disk reaching to 4 times its optical radius showing
evidence of recent star formation. The observation of this galaxy was
performed on December 9th 2007 with one single pointing in dithering mode
covering the optical radius of NGC\,4625. The spectroscopic data for this
object consists of 993 individual spectra.

\subsection{NGC 5474}

NGC\,5474 is a strongly lopsided spiral galaxy covering an area of
4.8\,$\times$\,4.3 arcmin in Ursa Major, at a distance of 7 Mpc. We observed this galaxy
with a standard mosaic configuration consisting of one central position and one
concentric ring. All pointings were observed in dithering mode. Observations
were carried out in two different periods; 4 positions were observed during
August 9th and 10th 2008, while 2 additional pointings were observed on the
27th April 2009. Given the distorted morphology of this galaxy, the central
position of the mosaic does not coincide with the bright bulge; a 30 arcsec offset in
declination was performed towards the south, so that the area covered by the
IFU mosaicking includes most of the optical disk of the galaxy in a symmetric
way. The area covered by all the observed positions is approximately 6
arcmin$^2$. The spectroscopic data for this galaxy totals 5958 individual
spectra.

\subsection{NGC 6643}

NGC\,6643 is a SAc galaxy in Draco, with a projected size of
3.8\,$\times$\,1.9 arcmin in the $B$-band, at a distance of 20 Mpc. 
A tailored mosaic pattern for this
galaxy was constructed in order to cover most of its optical area. 
Three pointings cover the surface of NGC\,6643 with a central position
centered on the bulge and two offsets of (37,\,34) and (--35,\,--34) arcsec in
(RA,Dec) in north-east and south-west directions respectively (see
\autoref{fig:sample}). Observations were performed on June 2nd 2008 for the
first 2 positions and on August 10th 2008 for the 3rd position, all of them in
dithering mode. At the time of the first observing run, we were able to
observe the supernova 2008bo, a SN type Ib located at 31'' north and 15'' west
of the nucleus of NGC\,6643. This galaxy has a complete spectroscopic mosaic
covering an area of approximately 2.8 arcmin$^2$. The data consists of 2979
individual spectra. However, due to a technical problem with the instrument
set-up, positions 1 and 2 do not cover the usual wavelength range, but are
shifted towards the red by approximately 100 \AA.

\subsection{NGC 6701}

NGC\,6701 is a small barred spiral in Draco with an angular size of
1.5\,$\times$\,1.3 arcmin, at a distance of 57 Mpc.
This galaxy was considered to be an isolated galaxy, but studies of
NGC\,6701 have discovered morphological and kinematical
features that are consequence of an interaction, most probably with a
companion at 73 kpc in projected distance \citep{Marquez:1996p3216}.
The observation of NGC\,6701 was carried out on August 9th 2008
with one single pointing in dithering mode covering the optical radius of the
galaxy (see \autoref{fig:sample}). The spectroscopic data of this galaxy
contains 993 individual spectra.

\subsection{NGC 7770 and NGC 7771}

The main target for this mosaic was the galaxy NGC\,7771, a barred spiral in
Pegasus with an optical $B$-band size of 2.5\,$\times$\,1.0 arcmin at a
distance of 59 Mpc. This galaxy is part of an interactive system containing
mainly the face-on spiral NGC\,7769 and the faint lenticular NGC\,7770
\citep{Nordgren:1997p3506}. The central part of
NGC\,7771 contains a massive circumnuclear starburst which was probably
triggered by the interaction with the other members of the group
\citep[][and references therein]{Smith:1999p3229}. Due to the projected size
of this galaxy, the mosaic pattern was constructed with three IFU positions. 
The central position of the mosaic has an offset of (--15,\,--15)
arcsec in (RA,\,Dec) from the geometrical centre of the galaxy (see
\autoref{fig:sample}). Two additional positions were observed with offsets of
(37,\,33) and (--37,\,--33) arcsec. A second member of the interacting group,
NGC\,7770, a small S0 galaxy with an optical size 0.8\,$\times$\,0.7
arcmin was observed within the field of the mosaic pattern. Observations
of all 3 positions were performed on October 30th 2008. The spectroscopic data
for this galaxy contains 993 individual spectra.

\subsection{Stephan's Quintet}

The Stephan's Quintet is one of the most famous and well-studied group of
galaxies, consisting of NGC\,7317, 7318A, 7318B, 7319 and
7320 in Pegasus. The distance to this compact group of galaxies has been in
debate due to the presence of the brightest member, NGC\,7320, which exhibits a
smaller redshift than the others, suggesting that is a foreground
object lying along the line of sight of the other four interacting galaxies.
Although some controversy prevailed
\citep{Balkowski:1974p3239,Kent:1981p3238}, recent observations by HST show
that individual stars, clusters, and nebulae are clearly seen in NGC\,7320 and
not in any of the other galaxies \citep[][and references
therein]{Gallagher:2001p3240,Appleton:2006p3237}. Four individual
pointings in non-dithering mode were observed on August 10th 2008, three of
which were centered at the bright bulges of NGC\,7317, 7319 and 7320, while
the last pointing was centered in configuration to cover NGC\,7318A and
NGC\,7318B (see \autoref{fig:sample}). The spectroscopic data for all
pointings of the Stephan's Quintet contains 1324 individual spectra.

\section{Data reduction}
\label{sec:reduction}

The reduction of IFS observations possesses an intrinsic complexity given the
nature of the data and the vast amount of information recorded in a 
single observation. This complexity is increased if one considers
creating an IFU spectroscopic mosaic of a given object for which the
observations were performed not only on different nights, but even in
different years, with dissimilar atmospheric conditions, and slightly
differing instrument configurations.

In this section we give an overview of the IFS data reduction process for
all the observations of the PINGS sample. In general, the reduction process
for the all pointings follows the standard steps for fibre-based integral
field spectroscopy. However, the construction of the mosaics out of the
individual pointings requires further and more complicated reduction steps
than for a single, standard IFU observation.
These extra steps arise due to the special mosaicking pattern for some of the
objects, the differences in the atmospheric transparence and extinction,
slight geometrical misalignments, sky-level variations, differential
atmospheric refraction, etc.
A complete explanation of the complex data processing for the creation of the
PINGS mosaics sample is beyond the scope of this paper, but the reader will
find a detailed description of the IFS data reducing in
\citealt{Sanchez:2006p331} (hereafter San06) and
additional information on the mosaicking technique for the PINGS sample in the
description of the PPAK-IFS survey of NGC\,628 (\citetalias{paperII}).

Following \citetalias{Sanchez:2006p331}, all the data reduction steps can be
summarised as follows: a) Pre-reduction. b) Identification of the location of
the spectra on the detector. c) Extraction of each individual spectrum. d)
Distortion correction of the extracted spectra. e) Application of wavelength
solution. f) Fibre-to-fibre transmission correction. g) Flux calibration. h)
Allocation of the spectra to the sky position. i) Cube and/or dithered
reconstruction (if any).

The raw data extracted from a PINGS observation consists of a collection of
spectra, stored as 2D frames, aligned along the dispersion axis.
The pre-reduction of the IFS data consists of all the corrections applied to the CCD
that are common to the reduction of any CCD-based data, i.e. bias
subtraction, flat fielding (in the case of PINGS, using twilight sky
exposures), combinations of different exposures of the same
pointing and cosmic ray rejection. The pre-reduction processing was performed
using standard IRAF\footnote{IRAF is
  distributed by the National Optical Astronomy Observatories, which are
  operated by the Association of Universities for Research in Astronomy, Inc.,
  under cooperative agreement with the National Science Foundation.} packages for
CCD pre-reduction steps while the main reduction was performed using the R3D
software for fibre-fed and integral-field spectroscopy data
\citepalias{Sanchez:2006p331} in combination with the E3D visualization software
\citep{Sanchez:2004p2632}.

On a raw data frame, each spectrum is spread over a certain number of pixels along
the cross-dispersion axis. The spectra are generally
not perfectly aligned along the dispersion axis due to the configuration of
the instrument, the optical distortions, the instrument focus and the
mechanical flexures. Therefore, in order to find the location of each spectrum
at each wavelength along the CCD and to extract its
corresponding flux, we made use of continuum illuminated exposures taken at
each pointing corresponding to a different orientation of the telescope. Each spectrum was
extracted from the science frames by co-adding the flux within an aperture of
5 pixels assuming a cut across the cross-dispersion axis found by iterative
Gaussian fits (see \autoref{sec:cross} for a detailed description of this
reduction step). 
Since the misalignments of the fibres with the pseudo-slit also affect the
wavelength solution, we require lamp exposures obtained at each observed
position to find a wavelength solution for each individual spectrum. 
Wavelength calibration was performed using HeHgCd+ThAr arc lamps obtained
through the instrument calibration fibres. Differences in the fibre-to-fibre
transmission throughput were corrected by creating a master fibreflat from
twilight skyflat exposures taken in every run.

The reduced IFS data can be stored using different data formats, all of which should
allow to store the spectral information in association with the 2D position on
the sky. Two data formats are widely used in
the IFS community: datacubes (3-dimensional images) and Row-Staked-Spectra
(RSS) files. Datacubes are only valid to store reduced data from instruments
that sample the sky-plane in a regular-grid or for interpolated data. In
this case the data are stored in a 3-dimensional FITS image, with two spatial dimensions and one
corresponding to the dispersion axis. RSS format is a 2D FITS image where the
{\it X} and {\it Y} axes contain the spectral and spatial information
respectively, regardless of their position in the sky. This format requires an
additional file (either a FITS or ASCII table), where the position of the
different spatial elements on the sky is stored. RSS is widely used by IFUs
with a discontinuous sampling of the sky, as it is the case for PPAK. We chose
to store the PINGS data in the RSS format, with corresponding position
tables.

Once the spectra are extracted, corrected for distortions, wavelength
calibrated, and corrected for differences in transmission fibre-to-fibre, they
must be sky-subtracted and flux-calibrated.
One of the most difficult steps in the data reduction is the correct
subtraction of the night sky emission spectrum. In long-slit spectroscopy the
sky is sampled in different regions of the slit and a median sky spectrum is
obtained by spectral averaging or interpolation. This is possible due to the
size of the long-slit compared with the size of the astronomical objects of
interest. However, in IFS the techniques vary depending on the geometry of
the observed object and on the variation of the sky level for a given
observation. By construction, many of the positions of the PINGS
mosaics (specially those in the galaxy centre) would fill the entire FOV of
the IFU, and none of the spaxels\footnote{Definition of IFS discrete spatial
  elements} would be completely free of galaxy ``contamination''.
In this case, we obtained supplementary sky exposures (immediately after the science
frames) applying large offsets from the observed positions and between
different exposures, we then used these ``sky-frames'' to perform
a direct sky subtraction of the reduced spectra. On the other hand, if the FOV is
not entirely filled by the object, it is possible to select those spaxels
(i.e. the sky-fibres in the case of PPAK) with spectra free of contamination
from objects, average them and subtract the resulting sky-spectrum from the
science spectra. We used this technique for observations in the last ring of a
mosaic or at the edges of the optical surface of the galaxies, where the
sky-fibres bundles did actually sample the sky emission.

Once the sky emission has been subtracted, we need to flux calibrate the
observed frames. Absolute spectrophotometry with fibre-fed spectrographs is
rather complex; as in slit-spectroscopy, where slit losses impose severe
limitations, IFU spectrographs can suffer important light losses due to the
geometry of the fibre-arrays. The flux calibration requires the observation of
spectrophotometric standard stars during the night.
Given that the PPAK IFU bundle does not cover the entire FOV due to gaps among
the fibres, the observation of calibration standard stars is prone to flux losses,
especially when the standards are not completely well centered in a single IFU
spaxel.
However we developed a method which takes into account the
flux losses due to the gaps in the fibre-bundle, the pointing misalignments
and PSF of the observed standard stars, as well as corrections for minor
cross-talk effects, airmass, local optical extinction and additional information provided
by broad and narrow-band imaging photometry in order to obtain the most
accurate possible spectrophotometric flux-calibration within the limits
imposed by the instrumentation.

\begin{figure}
  \includegraphics[width=0.49\textwidth]{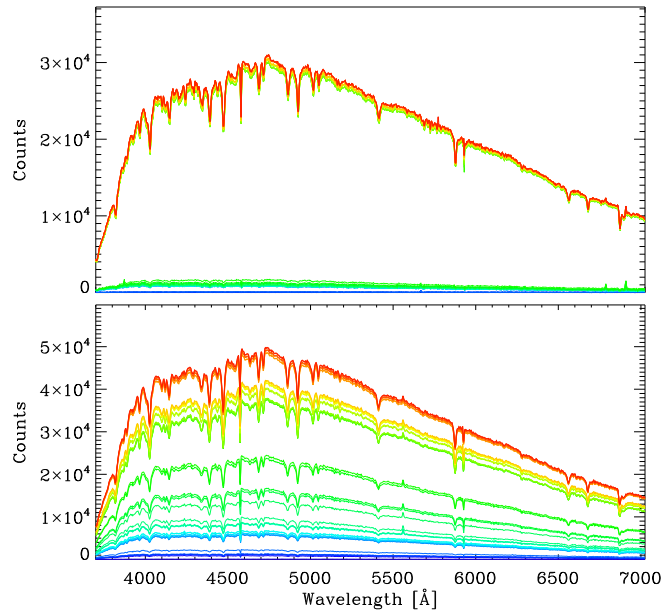}
  \caption{Effect of standard star flux loss due to bad seeing and
    misalignments. On the top panel, the standard star BD+25d4655 was observed
    during good atmospheric conditions and was well centred on the
    central fibre of the PPAK bundle. The figure contains 37 spectra
    corresponding to the 36 concentric fibres plus the central one. Most of
    the flux falls in the central fibres (red colour spectra in the online version),
    while very small residuals are seen in the
    rest of them (green-blue colour spectra). The bottom panel shows the
    spectra of the same star observed
    during turbulent atmospheric conditions and when the star was not well
    centred in the fibre bundle. In this case the flux is spread over a large
    number of concentric fibres and the individual contribution of each of
    them is important for the total observed flux. 
}
  \label{fig:std_fibers}
\end{figure}

A total of six standard stars from the Oke spectrophotometric candles
\citep{Oke:1990p2634} were observed for the purpose of flux calibration during
the observing runs. These frames were reduced following the basic procedure
described above. To counteract the loss of flux, the observed spectrum of a
standard star was obtained by adding up the spectra from consecutive
concentric spaxel rings centered on the fibre where most of the standard's
flux is found, until a convergence limit was found (see
\autoref{fig:std_fibers}). We then determined a night
sensitivity curve as a function of wavelength by comparing the observed flux
with the calibrated spectrophotometric standard spectrum considering the
filling factor of the fibre-bundle. We applied this function to the science
frames considering corrections for the airmass and the optical extinction due
to the atmosphere as a function of wavelength to get relative flux-calibrated
spectra for each position.

\autoref{fig:all_std} shows the shape and relative magnitude of nine
sensitivity curves obtained during the three years of observations. 
The difference in the vertical scale reflects the variation of the
spectrophotometric transmission during different nights and observing
runs (the flux calibration obtained by applying these sensitivity curves is
just a relative one, an absolute calibration is obtained by re-scaling by a
factor derived after the comparison with the broad-band imaging by applying
the method described below).
However, the differences on the shape from one sensitivity curve to another as
a function of wavelength, reflect the intrinsic dispersion of the flux
calibration. 
The bottom panel of \autoref{fig:all_std} shows the variation of the
sensitivity curves as a function of wavelength after a grey shift with respect
to an arbitrary spline fitting normalised at the wavelength of H$\beta$
(4861\,\AA). This normalisation wavelength was chosen as most spectroscopic
studies normalise the observed emission line intensities to the flux in
H$\beta$. The maximum variation from the blue end of the spectrum compared to
the red one is of the order of $\sim$ 0.15 mag, corresponding to a maximum calibration
error of $\sim$ 15\% due solely to the intrinsic dispersion in the relative
flux calibration. However, the actual RMS is less than 0.1 mag, corresponding
to a typical error in the relative flux calibration of less than 10\%.

\begin{figure}
  \includegraphics[width=0.49\textwidth]{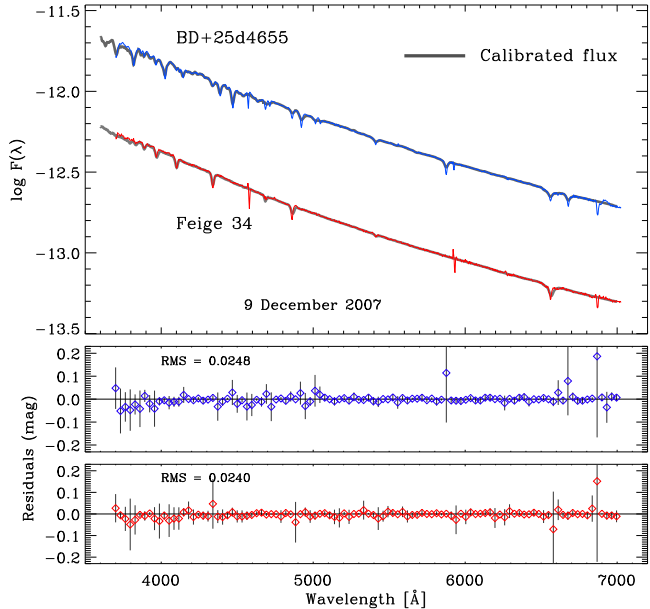}
  \caption{Observed standard stars on the 9th December 2007. The top panel
    figure shows a comparison of the night flux calibration applied to the
    observed stars (in colour in the online version) and the corresponding
    calibrated fluxes. The
    panel below show the residuals in magnitudes as a function of
    wavelength for each star. The points with relatively large
    deviations are due to strong sky emission lines and cosmetic defects of
    the CCD.}
  \label{fig:flux_comp}
\end{figure}

For those galaxies with suitable multi-band photometric data available in the
literature (e.g. NGC\,628, NGC\,3184, NGC\,4625, NGC\,5474), we used the
first-order calibration of either the central pointing of a mosaic or the
position with the highest S/N and best sky-subtraction to perform an
additional correction to the absolute flux calibration by comparing the
convolved spectra observed in this field (taking into account fibre apertures
and filters' response functions) to the corresponding flux measured by $B$, $V$, $R$
broad-band and H$\alpha$ narrow-band imaging photometry for the same
position. This procedure ensures a very precise flux calibration and sky
extinction correction for this master pointing.
To our knowledge, no other IFS observations have ever attempted to get such
(instrument-limited) spectrophotometry accuracy. All galaxies belonging to
the SINGS sample were corrected by this method using their ancillary data as
described in more detail in \citetalias{paperII}.

After reducing each individual pointing with a first-order flux calibration
and with the help of the absolute flux-calibrated master pointing, we built a
single RSS file for the whole mosaic following an iterative procedure. 
The process starts in the master pointing chosen for a specific mosaic,
i.e. the pointing that has the best possible flux calibration and sky
extinction correction, with the best signal-to-noise and the most optimal
observing conditions regardless of the geometric position of the pointing in
the mosaic. Taking this master pointing as a reference, the mosaic is
constructed by adding consecutive pointings following the particular mosaic
geometry. During this process, the new added pointing is re-scaled by using
the average ratio of the brightest emission lines found in the overlapping
spectra (which is then replaced by the average between the previous pointing
and the new re-scaled spectra). In most cases the scale factor is found to be
between 0.7 and 1.3 with respect to the master pointing. 

However, this ratio is wavelength dependent (specially in the cases of
variable photometric conditions between the observations). Therefore as a
second-order correction, we fitted the variations found between the previous
pointing and the new re-scaled overlapping spectra to a low order polynomial
function and divided all the spectra in the new pointing by the resulting
wavelength dependent scale. This correction has little effect on the data when
the observations were performed during the same or consecutive nights, as it is
the case for the small mosaics. However, we accounted variations after all the
possible corrections of the order of 10-15\% in the extreme cases when the
observations were carried out at different epochs (e.g. NGC\,628, NGC\,3184). 
This level of error is what we expect from observations performed during
different nights and observing runs, reflecting the variation of
the spectrophotometric transmission from night to night (see
\autoref{fig:all_std}).
The procedure was repeated for each mosaic until the last pointing is included
(except for the Stephan's Quintet, where not actual overlapping occurs),
ending with a final set of individual spectra and their corresponding position
tables. This process ensures a homogenous flux calibration and sky extinction
correction for the entire data set.

\begin{figure}
  \includegraphics[width=0.49\textwidth]{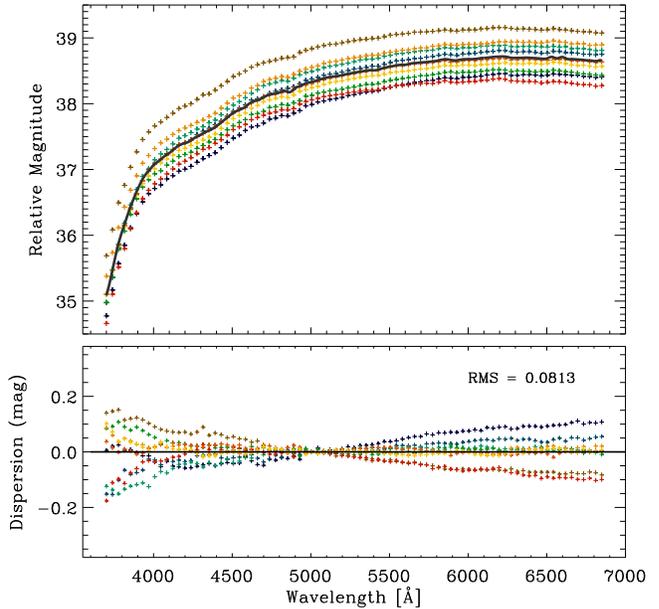}
  \caption{Variation of the spectrophotometric transmission as a function of wavelength.
    The top panel shows nine sensitivity transmission curves using different
    calibration standard stars. The thick dark line is a 3rd degree spline
    fitting to the data. The panel below shows the dispersion in
    magnitudes after a grey re-scaling of all the sensitivity curves with respect to
    the spline fit normalised at the H$\beta$ wavelength (4861\,\AA).}
  \label{fig:all_std}
\end{figure}

\section{Errors and uncertainties in the data sample}
\label{sec:errors}

During the process of data reduction and basic analysis, we have identified
several possible sources of errors and uncertainties in the PINGS data
set. Each of them contribute in a different way and 
magnitude to the overall error budget associated with the observations. 
These are in order of importance: 1) sky subtraction, 2) flux calibration, 3)
differential atmospheric refraction (DAR), 4) cross-talk, and 5) second order spectra.
In this section we describe the nature of each of these sources of errors, the
tests performed in order to understand their effects on the accuracy of the
data, and the techniques applied to minimise them.

\subsection{Sky subtraction}
\label{sec:sky_sub}

As mentioned before, sky subtraction is one of the most
difficult steps in the IFS reduction process and it is particularly complex
for the nature of the observing technique of PINGS. 
As described in \citet{Sanchez:2006p331}, a deficient sky subtraction in this sort of data has several
consequences: ``the contamination of the sky emission lines along the spectra
which prevents the detection and/or correct measurement of relatively weak
nebular emission lines (e.g. the weak temperature sensitive \oiii
\lam4363, which is located in the same spectral region as the
strong Hg I \lam4358 sky line), and also affects the shape and
intensity of the continuum, which is important for the analysis of the stellar
populations and the determination of reddening''.
In fact, we made use of the mosaicking method in order to find the best
possible sky subtraction. Due to the shape of the PPAK bundle and by
construction of the mosaics in the standard mosaic
configuration, 11 spectra of a given pointing (corresponding to one edge of the
hexagon) overlap with the same number of spectra from the previous
pointing (see NGC\,628 or NGC\,3184 in \autoref{fig:maps_1}). This allows the
comparison of the same observed regions at different times and with different
atmospheric conditions. 

For a non-standard configuration the number of overlapping fibres
is larger (e.g. NGC\,3310). These overlapping spectra can be compared and used
to correct for the sky emission of the adjacent frame. However, prior to
performing the sky subtraction it is required to visually check that
no residual of the galaxy is kept in the derived spectrum. This can be the
case if the transmission changed substantially during the observation of the
adjacent frames. These techniques proved to result in good sky subtraction in
most cases. On the other hand, when we were forced to obtain supplementary
large-angle offset sky-exposures for the inner pointings in the mosaics that
were completely filled by the target, we found that when the sky exposure is
taken within a few minutes of the science exposure it produces a good subtraction.
For those cases in which the atmospheric conditions changed drastically and/or
the sky subtraction appeared to be poor, we combined different sky frames
with different weights to derive a better result.

One way of assessing the goodness of the sky subtraction is to check for sky
residuals in the subtracted spectra.
The galaxy mosaic more prone to be affected by residuals in the sky
subtraction is NGC\,628, which as explained in \autoref{sec:obs}, was observed
during six nights along four observing runs.
Therefore we would expect that the spectroscopic mosaic of this galaxy would
show the most extreme effects due to the sky subtraction to be found in the
PINGS sample, given all the variations in transparency and photometric
conditions of the night-sky along the three years of observations.

In order to obtain a quantitative assessment of the quality of the sky
subtraction, we performed two different data reductions of the spectroscopic
mosaic of NGC\,628. In the first reduction, the sky subtraction was performed
directly with the average spectrum of the sky fibres at each position, without
considering the overlapping spectra between pointings, and not accounting for
the object contamination in the sky fibres. Therefore, in this first reduction
we applied a ``poor'' sky subtraction. For the second reduction, we applied an
individual sky subtraction per mosaic position using the techniques described
above, i.e. applying corrections using the overlapping spectra,
checking for galaxy residuals in the derived sky spectrum, using the sky
exposures obtained by large offsets for the most internal regions of the
galaxy, and combining different sky frames with different weights in those
cases when there were important changes in the transmission between pointings
observed during the same night. We refer to this reduction as the ``refined''
sky subtraction.

Airglow is the most important component of the light of the night-sky spectrum
at Calar Alto observatory, although a substantial fraction of the spectral
features is due to air pollution \citep{Sanchez:2007p1276}.
The strongest sky line in the Calar Alto night-sky spectrum is the \oi
\lam5577 line, followed by the \oi \lam6300 line, both produced by
airglow with a notorious stronger effect near twilight.
A deficient sky subtraction can be recognised by residual features of the sky
lines in the derived spectra, this effect is clearly seen in the \oi
\lam5577 sky line which is located in a spectral region without any
important nebular emission line. In general terms, (without considering
variations in the transparency of the sky), a residual in emission of this
line would imply a subtraction of the sky spectrum of slightly lower strength
than required, while an absorption feature would imply an over-correction.

In order to make a comparative analysis of the strength of the sky residuals
in the two data reductions of NGC\,628 described above, we measured the
equivalent width (EW) of the residual features centered at the \oi \lam5577 line.
For numerical reasons (regions of null continuum), the local continuum in the
neighbourhood of the \oi \lam5577 sky line was re-scaled to a flux level of
10$^{-16}$\,erg\,s$^{-1}$\,cm$^{-2}$\,\AA$^{-1}$ in every single spectrum of
both mosaics. EWs with negative sign correspond to residual emission features,
while positive EWs correspond to absorption features.

\begin{figure}
  \includegraphics[width=0.49\textwidth]{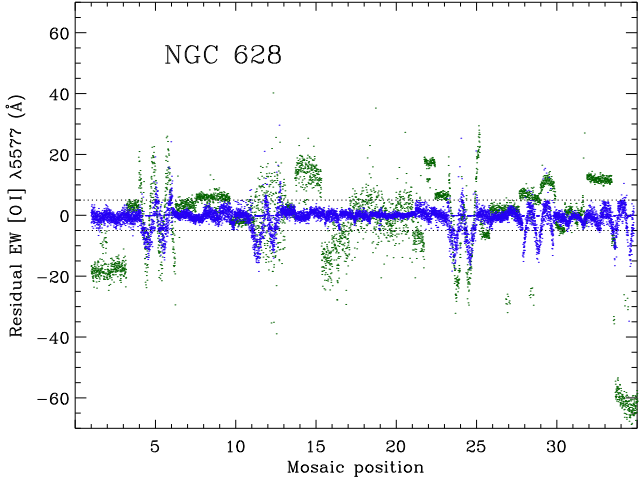}
  \caption{Equivalent width values for the emission and absorption residual
    features of the [O\,{\scriptsize I}] \lam5577 sky line as a function of the observed
    position for the spectroscopic mosaic of NGC\,628. Positive values
    correspond to absorption features, while negative values to emission
    residuals. The local continuum level was re-scaled to the same value in order
    to make this comparison. The green dots (in the online version) correspond
    to a {\em poor} sky subtraction, while the blue dots represent the
    {\em refined} reduction as explained in the text. The two horizontal dotted
    lines mark the threshold EW values of residuals features corresponding to a
    good sky subtraction.}
  \label{fig:sky_resid}
\end{figure}

\begin{figure}
  \includegraphics[width=0.49\textwidth]{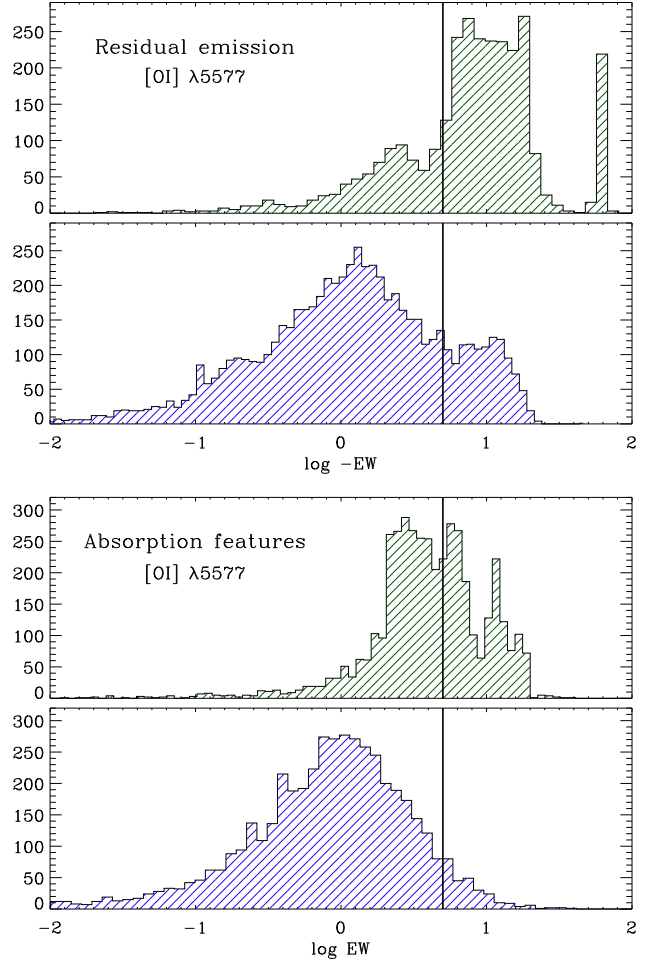}
  \caption{Histograms of the EW residual features shown in
    \autoref{fig:sky_resid}. The top panels shows the distribution of EW
    values for the emission residuals. The bottom panels shows the EW values
    for the absorption features. All EW values are shown in logarithmic scale
    (assuming a positive EW value for the emission residuals). The top
    histograms in both cases correspond to the {\em poor} sky subtraction, while
    the bottom histograms correspond to the {\em refined} reduction (green and
    blue colour histograms in the online version). The $\sim$\,5 \AA\ EW
    threshold value is shown as the vertical line in all the histograms, residual
    values to the right of this line can be considered a deficient sky
    subtraction as explained in the text.}
  \label{fig:sky_ew}
\end{figure}

\autoref{fig:sky_resid} shows the value of the EW residuals for the \oi
\lam5577 line for both data reductions as a function of the pointing
position in the spectroscopic mosaic. Each position {\em bin} contains 331
values corresponding to the number of spectra per pointing, a total of 11104
values are shown, corresponding to the 34 positions observed for NGC\,628.
The green dots correspond to the {\em poor} sky subtraction reduction, while the
blue dots correspond to the {\em refined} sky subtraction (see the online
version of this plot). 
There is a considerable amount of scatter of the EW residual value for the 
{\em poor} sky subtraction compared to the {\em refined} one. 
In the first two pointings of the mosaic (which correspond to central positions of the
galaxy), there are strong residuals in emission for the {\em poor} reduction,
while the residuals have been minimised in the {\em refined}
one. \autoref{fig:sky_resid} shows clear evidence of those pointings in
which the sky transparency varied by a considerable amount (positions 4, 5,
11, 12, 23, 24, 28, 29, 33, 34). In all the pointings, the scatter in the
residuals is improved in the {\em refined} reduction with respect to the first
one. This effect is more notorious between positions 13 to 22. The {\em poor}
sky subtraction yields very strong sky residuals in emission for positions 33
and 34, while in the {\em refined} reduction these are minimised.

At the chosen continuum level used for this exercise, a (absolute) value of 5
\AA\ in EW for the \oi \lam5577 residual line in emission
corresponds approximately to a flux intensity value of 5 $\times$
10$^{-16}$\,erg\,s$^{-1}$\,cm$^{-2}$, while a value of 10 \AA\ corresponds to
$\sim$ 10 $\times$ 10$^{-16}$\,erg\,s$^{-1}$\,cm$^{-2}$. The average flux
intensity of the \oi \lam5577 sky line in Calar Alto is of the order of 33 $\times$
10$^{-16}$\,erg\,s$^{-1}$\,cm$^{-2}$ \citep{Sanchez:2007p1276}. However, from
a sample of 500 sky spectra acquired during the three years of observation we
measured the intensity of the \oi \lam5577 in the range between 30 and 60
$\times$ 10$^{-16}$\,erg\,s$^{-1}$\,cm$^{-2}$, with a mean value of 44. Therefore, a
value of 5 \AA\ in EW for a residual emission feature would correspond to
$\sim$\,8--10\% of the total emission of the \oi \lam5577 line. Visual
inspection of the spectra with emission residual of the order of 5 \AA\ in EW
confirms that this value could be considered as the threshold for a good
sky subtraction. Spectra with emission or absorption residuals with absolute
EW values less than 5 \AA\ could be considered to have a good subtraction, for
features above this value the effects of a deficient sky subtraction are
evident.

The two horizontal dotted lines in \autoref{fig:sky_resid} indicate the $\pm$5
EW threshold value for both emission and absorption features. These two lines
encompass a region for which the spectra can be considered with a good sky
subtraction. The {\em poor} sky subtraction (green) shows a lot of scatter
and a small fraction of the spectra falls within these limits. On the other
hand, for the {\em refined} reduction (blue) a total of 9629 spectra fall within
these limits, i.e. 87\% of the total mosaic. The number of spectra with sky
subtraction problems for which $|$EW$|$ $>$ 5 \AA\ is 1475, i.e 13\% of the
mosaic, these spectra are found in those pointings with extreme transparency
variations as expected.

\autoref{fig:sky_ew} shows the histograms of the EW values for
both data reductions, the {\em poor} sky subtraction in green and the
{\em refined} reduction in blue colour following \autoref{fig:sky_resid} (see
the online version of this plot). The top panels show the
distribution of residual emission values, while the bottom panels show the
absorption residual features for the \oi \lam5577.
The $\sim$\,5 \AA\ EW threshold value is shown as the vertical line in the
histograms, residual values to the right of this line can be considered a
deficient sky subtraction.
Visual inspection of the spectra shows that, at the continuum level used for
this comparison, emission or absorption features with values of log($|$EW$|$)
$\le$ 0 could be considered negligible and within the statistical noise of the
spectra. The residual emission histograms show that the {\em poor} sky
subtraction produces a large number of strong residuals with values of $|$EW$|$
$>$ 5 \AA, even reaching $|$EW$|$ $\sim$ 60 \AA. The majority of the residual
values in {\em refined} sky subtraction are found at log($|$EW$|$) $\sim$ 0,
corresponding to negligible residual values, however there is a small tail of
strong emission residuals for which $|$EW$|$ $>$ 5 \AA\ ($\sim$\,18\% of the
total emission residuals).
The distribution of EW values of the absorption features for the {\em poor}
subtraction is approximately centered at the threshold limit, while for the {\em refined}
reduction, the values are nearly normally distributed with a centre value of
log(EW) $\sim$ 0 with a small tail of strong absorption values ($\sim$\,7\%)
due most likely to an over subtraction of the sky spectrum.

The {\em refined} sky subtraction was the final adopted one for the
spectroscopic mosaic of NGC\,628. All the sky subtraction techniques implemented
showed that the quality of the derived spectra was improved by a considerable
amount compared to a standard sky subtraction. Most of the sky residuals are
within the limits of a reasonably good sky subtraction. The spectra with
strong features are found for those positions in which the photometric
conditions changed drastically during the night or observing run. This
residual analysis allows to identify those pointings with strong sky
variations and thus, to flag the spectral data for future analysis. The sky
subtraction for the rest of the PINGS sample was performed similarly to the
{\em refined} technique described above. Therefore we applied the best possible
sky subtraction to all the spectroscopic mosaics  within the limitations
imposed by the IFS data itself.

\subsection{Detection of the [O\,III] $\lambda$4363 line}
\label{sec:simu}

The Hg \lam4358 sky line strongly affects any attempt to measure precisely the
emission of the faint temperature-sensitive \oiii \lam4363 line in any object
with a low redshift.
The fact that the strength of this line decreases with increasing abundance
\citep{Bresolin:2006p222}, in combination with typically faint \hh regions and
low spectroscopic resolution limits importantly the detectability of this key
diagnostic line.

In order to assess the significance of the detection of the \oiii
\lam4363 given the contamination of the Hg \lam4358 sky line in our data, we
performed a simulation of the detectability of the \oiii \lam4363 line for
a given range of redshifts and line intensity strengths. We simulated a pure
emission line spectrum including the H$\gamma$ \lam4340 and \oiii
\lam4363 lines at the same spectral resolution of the PINGS
observations; we assumed a normally distributed I(\lam4363)/H$\gamma$
ratio with a mean value of 0.10\,$\pm$\,0.05, corresponding to typical
values found in previous spectroscopic studies where the \oiii \lam4363
line was detected in \hh regions within the metallicity range of
the PINGS sample \citep[e.g.][]{McCall:1985p1243};
we did not consider higher ratios ($\sim$ 0.25\,$\pm$\,0.10) which are
representative of extremely low metallicity objects
\citep[e.g.][]{Pagel:1992p520,Izotov:2004p352}.
We added a random statistical noise of 0.05 RMS at the continuum level
constructed from the observed spectroscopic data. A sample of 540 sky
spectra were selected among all the observing runs during the three years of
observations (considering very different photometric conditions). A flux
calibrated sky spectrum was created out of these selected spectra.
This sky spectrum was added to the
previous emission line plus the noise described above to create a
simulated ``observed'' spectrum. 
An average sky spectrum constructed from a random subset of 36 sky spectra
(the number of PPAK sky-fibres) was then subtracted from the simulated ``observed''
spectrum to obtain a ``sky-free'' spectrum.

\begin{figure}
  \includegraphics[width=0.49\textwidth]{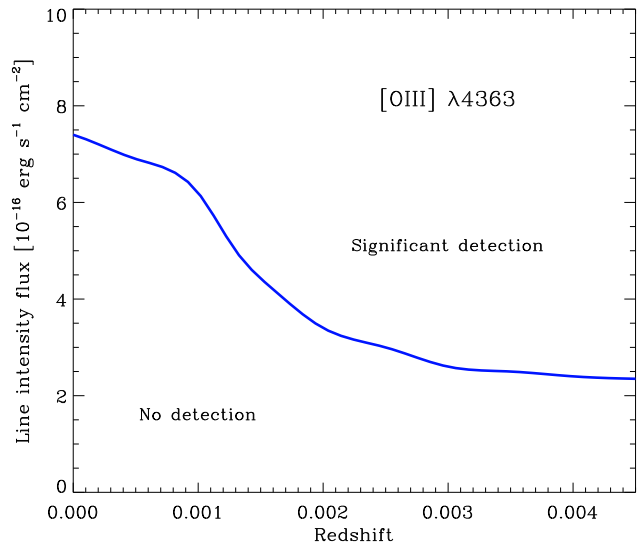}
  \caption{Detectability of the [O\,{\small III}] \lam4363 line as a function of
    redshift and line strength according to the simulation described in
    \autoref{sec:simu}. For a given redshift, the [O\,{\small III}]
    \lam4363 can be significantly detected for flux values above the
    thick line, which marks the region at which the difference between the
    observed and simulated line intensity is of the order of 15\%. The
    contamination effect of the Hg \lam4358 disappears for redshift
    values larger than 0.004.}
  \label{fig:simu}
\end{figure}

Emission line intensities were then measured simultaneously for both lines in
the simulated ``sky-free'' spectrum using the techniques described in
\autoref{sec:emission}. These line intensities were then compared with the
flux of the pure emission lines. For a given redshift, we varied the emission
line strengths of the simulated spectrum from high to lower values
until the significance of the detection of the \oiii \lam4363 fell
drastically. We performed 500 realisations of the emission line intensity
measurements for a given redshift and for a given line strength.
\autoref{fig:simu} shows the results of the simulation, the thick
line represents the region at which the difference between the line intensity
measured from the simulated ``sky-free'' spectrum and the flux from the pure
emission line is of the order of 15\%. According to the simulation, observed flux
values of the \oiii \lam4363 above this line can be significantly
detected at a given redshift. For flux values below this region the
significance of the detection is negligible as it is mostly affected by the
statistical noise of the data. The contamination effect of the Hg \lam4358
disappears for redshift values larger than $\sim$ 0.004, where the
detectability of the \oiii \lam4363 depends on the signal-to-noise of the
spectrum at low line-intensity levels.
Experience with the data has proven that the simulation described in this
section places very good limits on the detectability and potential measurement
of this faint line, although after detection, individual and visual
inspection of the spectra has to follow in order to correctly assess the
usefulness of this line.

\subsection{Flux calibration}
\label{sec:flux_calib}

Several refinements in the observation technique and standard flux recovery
were applied to the pipeline which improved substantially the accuracy of the
sensitivity functions obtained after every standard candle observation. During
most of the observing runs we observed different standard stars per night at
different airmasses in order to asses the variation in transmission and
its effect on the relative flux calibration.

We generated several sensitivity curves following the standard pipeline
procedure in R3D, changing the key parameters that could affect the accuracy
of the derived sensitivity function (e.g. order and type of the fitting
function, extinction, airmass, smoothing, etc.). Furthermore, we made a
comparison of the response curves obtained through R3D and the ones obtained
using standard long-slit flux calibration routines in IRAF after performing
all the appropriate corrections and transformations for the two different
kinds of data. We even derived whole-run sensitivity curves after the
combination of several response curves for a given observing run and applied
the derived calibration to the observed standard candles as a proof of
self-consistency. In all cases, we found very consistent results in the final
relative flux calibration. As described in \autoref{sec:reduction}, even
without a re-calibration using broad-band imaging, the spectral shape and
features are reproduced within the expected errors for an IFS observation
($\sim$ 20\% in the absolute sense) along the whole spectral range, with a
small increase in the blue region ($\lambda <$ 3800 \AA) due in part to the
degradation of the CCD image quality and instrumental low sensitivity towards
the blue ($\sim$ 2\,--\,5\%, telescope/atmosphere excluded) in this spectral
region \citetext{Roth, priv.\ comm.}.

For those galaxies with available multi-band photometric data, small
differences in the transmission curves of the filters and astrometric
errors of the built mosaics bring some uncertainties in the derived flux
ratios that contribute to the overall standard deviation when applying the
photometric re-calibration described before.
The errors in the first case are difficult to estimate, however the latter
ones were estimated by simulating different mosaic patterns by applying 
normally distributed random offsets of the simulated fibre-apertures with mean
values of 0.3, 0.5 and 1.3 arcsec on the broad-band images and then comparing
the extracted spectrophotometry.
From the results of these simulations and considering that the IFS mosaics
were re-centered using the information directly from the aperture photometry,
we expect that the location of the fibres lies within 0.5 arcsec, and
therefore the error due to the uncertainty in the astrometry would be of the
order of 10\%.
Based on these results, we estimate a spectrophotometry accuracy better than
$\sim$\,0.2 mag, down to a flux limit corresponding to a surface brightness of
$\sim$\,22 mag/arcsec$^2$ \citepalias[see][]{paperII} when we apply the re-calibration
derived by the flux ratio analysis.
In the following section we compare our data with previously published
spectrophotometrically calibrated data. The spectral shape, the
spectral features and emission line intensities match remarkably well, even
for those objects for which no re-calibration was performed (see \autoref{fig:integ}).
As stated above, to our knowledge no other IFU observation has ever
implemented such corrections in order to obtain an instrumental-limited
spectrophotometry accuracy as in this work.

\subsection{Differential Atmospheric Refraction}
\label{sec:dar}

An important systematic effect in any spectroscopic observation is due to
the refraction induced by the atmosphere, which tends to alter the apparent
position of the sources observed at different wavelengths. By definition, there
is no refraction when the telescope is pointed at the zenith, but for larger
zenith angles the effect becomes increasingly significant.
For IFU observations, this has the consequence that, when comparing for
example the intensities at two different wavelengths (e.g  the emission line
ratio of a source), one will actually compare different regions, given that
different wavelengths are shifted relative to each other on the surface of the
IFU. In theory, one is capable of performing a correction of DAR for a given
pointing without requiring knowledge of the original orientation of the
instrument and without the need of a compensator, as explained by
\citet{Arribas:1999p2670}.
The correction of DAR is important for the proper combination of different IFS
exposures of the same object taken at different altitudes and under different
atmospheric conditions, and for the proper alignment of a mosaic and dithered
exposures, as it is the case of most PINGS observations.
An IFS observation can be understood as a set of narrow-band images with a
band-width equal to the spectral resolution \citepalias{Sanchez:2006p331}. These
images can be recentered using the theoretical offsets determined by the DAR formulae
\citep{Filippenko:1982p2664} by tracing the intensity peak of a reference
object in the FOV along the spectral range, and recentering it. 
The application of this method is basically unfeasible in slit spectroscopy,
which represents one additional advantage of IFS.
The correction can be applied by determining the centroid of a particular object or
source in the image slice extracted at each wavelength from an interpolated
data cube. Then, it is possible to shift the full data cube to a common
reference by resampling and shifting each image slice at each wavelength
(using an interpolation scheme), and storing the result in a new data cube.
A pitfall of this methodology is that the DAR correction imposes always an
interpolation in the spatial direction as described above, a 3-dimensional
(3D) data cube has to be created for each observed position, reducing the
versatility given by the much simpler and handy RSS files.

We have to note here that the widely accepted formulation summarised by the
work of \citeauthor{Filippenko:1982p2664} and the concept of parallactic angle
are just a first order approximation to the problem.
All this theoretical body is based on the assumption that all different
atmospheric layers have an equal refraction index, are flat-parallel and 
perpendicular to the zenith. While this approximation is roughly valid, 
there are appreciable deviations due to the topography and landforms at the
location of the telescope, since they alter considerably the structure of 
the low-altitude atmospheric layers. Therefore, the ``a posteriori'' correction
of the DAR effect, only possible when using IFS, is the most accurate approach
to the problem.

In general, the effects due to DAR in IFS are only important for IFUs with small
spatial elements ($\le$\,1.5 arcsec) while for large ones (as it is the case of
PPAK), the effect is mostly negligible, as experience with the instruments
shows, especially when the airmass of the observations is 1.1 or below
\citep{Sandin:2008p2669}.
According to the DAR formulas one can calculate the angular separation in
arcsec due to this effect for two different wavelengths under typical
atmospheric conditions for a range of airmasses \citep[][Table
1]{Filippenko:1982p2664}. If we consider for example the wavelengths of
H$\alpha$ at \lam6563 and \oii \lam3727
emission lines, the angular separation due to DAR is smaller than the radius
of the PPAK fibres (1.35 arcsec) for airmasses below 1.3. 
Nevertheless, for each object in the sample we analysed those
pointings which were observed at an airmass $>$ 1.2 in order to test any effects due to
DAR in our data. We transformed these individual pointings into 3D data cubes
(with a scale of 1''/pixel)
and we selected suitable sources within the field (e.g. foreground stars,
compact bright emission line regions) to perform a DAR correction creating
continuum maps of these bright sources and looking for spatial deviations
along the dispersion axis. No significant intensity gradients were found in
any of the test fields.
Additionally, we looked for regions in which we could observe emission in the
blue (e.g. \oii \lam3727) but no emission in the red (e.g. H$\alpha$
\lam6563) and vice versa. In most of the cases we did not find strange
[O\,{\footnotesize II}]/H$\alpha$ ratios, although for some pointings we did
not have enough bright \hh regions to perform this exercise. However, for a
number of pointings we did find peculiar deviations from the flux measured in
both part of the spectra. All these pointings were observed with an airmass
$>$ 1.4 and correspond to: NGC\,1637 (all 7 dithered pointings), NGC\,6701
(all 3 pointings), and NGC\,5474 (pointings 3, 4 \& 5).
These particular pointings could be individually converted to 3D data cubes
and be further analysed and corrected for any DAR effects. In the case of
particular studies which require a degree of spatial precision, one has to
bear in mind the significance of the spatial-spectral information derived from
the individual fibres of these pointings. Integrated spectra over a large
aperture ($\sim$ 5 arcsec) could be considered reliable.

Considering in general that the effects of DAR are very small given the
relatively large size of the spatial elements of PPAK, that $\sim$\,70\% of
the observations were performed with an airmass $\le$ 1.35, and that we found
no significant evidence of DAR effects in the analysis of our data (apart from
the flagged pointings described above), we decided that no DAR correction
should be applied when building the final spectroscopic mosaics of the PINGS
sample, avoiding the transformation of the RSS files to 3D data cubes and the
undesirable interpolation of data.
It is important to note that problems caused by atmospheric dispersion cannot
be completely avoided in the spectroscopic study of extended objects like any
other IFS observation, only for those observations performed in dithering mode
a correction for the effect of DAR can be sought.

\begin{figure}
  \includegraphics[width=0.49\textwidth]{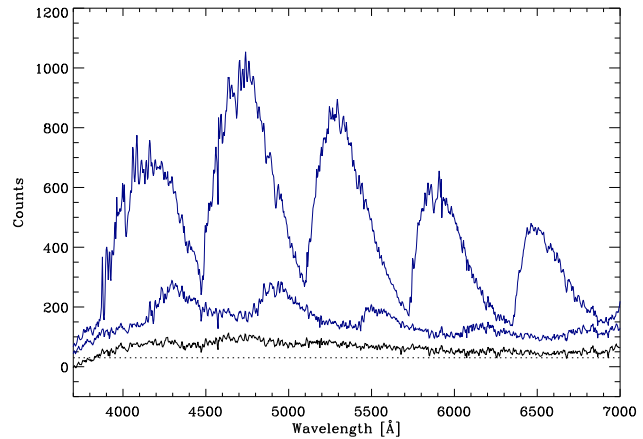}
  \caption{Example of a strong cross-talk effect in a fibre adjacent to a very
    bright source, in this case a standard star with a maximum of 40000
    counts. The top spectrum shows the full cross-talk
    effect when no correction is applied. The middle spectrum shows a
    first-order correction obtained by minimising the pixel extraction
    aperture (but losing flux in the process). The bottom spectrum shows the
    correction of the cross-talk by the improved {\em Gaussian-suppression} method
    (described in \citetalias{paperII}) with an extracted spectrum of the same
    order of intensity as the average background signal (dotted line) found in
    the nearby fibres for this particular observation. The apparent
    periodicity of the cross-talk with respect to the wavelength is just an
    effect of the spectra misalignments and distortions on the CCD.}
  \label{fig:cross}
\end{figure}

\subsection{Cross-talk}
\label{sec:cross}

One of the critical reduction steps is to extract the flux
corresponding to different spectra at each pixel along the dispersion
axis (see \autoref{sec:reduction}). However, this is not straightforward as
there could be contamination by flux coming from the adjacent spectra,
i.e. cross-talk. 
This contamination effect may produce a wrong interpretation of the data. For
observations performed with PPAK, and due to the geometrical construction of
the instrument, the adjacent spectra on the plane of the CCD
may not correspond to nearby locations in the sky plane
\citep{Kelz:2006p3341}. Furthermore, the position of the calibration fibres
(which are located in between the science ones along the pseudo-slit) also
contributes to overall contamination. Therefore, the cross-talk effect would
potentially mix up spectra from locations that may not be physically related
(because of the spectra position on the CCD) or from spectra with completely
different nature (i.e. calibration fibres).
Given that the cross-talk is an incoherent contamination it is preferable to keep it as low
as possible, an average value of $\sim$\,1\% with a maximum of $\sim$\,10\%
\citepalias[][and references therein]{Sanchez:2006p331} seems technically as a
good trade-off.

Given the limited size of CCDs and the need to record as many spectra as
possible within that reduced area, IFS observations will always face a certain
level of cross-talk. The FWHM of the projected profile along the
cross-dispersion axis is normally defined by the design of the spectrograph
and the size of the input fibres, placing a limit in the selected aperture. 
As explained by \citetalias{Sanchez:2006p331}, the PPAK
spectra profiles have a FWHM of the order of $\sim$\,2.3 pixels,
and $\Delta$\,peaks of $\sim$\,5 pixels in the 2\,$\times$\,2 binning mode (as
was the case for all PINGS observations). Selecting an aperture size of the
order of $\Delta$\,peaks seems to be an acceptable compromise between
maximising the recovered flux and minimising the cross-talk. 
Several methods have been implemented to minimise the effect of cross-talk, in
particular, \citetalias{Sanchez:2006p331} developed an efficient technique
named {\em Gaussian-suppression} that reduces the effects of the cross-talk and
maximises the recovered flux to within 10\% of the original values for any
spaxel at any wavelength. However, for certain raw frames which were too
crowded or when we targeted bright sources within the field (e.g. foreground
stars, galaxy bulges), we still found some level of contamination that could
not be considered negligible. 

Therefore, we improved the {\em Gaussian-suppression} technique to a new
method which increases the signal-to-noise ratio of the extracted spectra and
reduces the effect of the cross-talk compared to previous extractions. 
The new technique is explained in detail in \citetalias{paperII}, and it is
now fully implemented in the standard R3D package
\citepalias{Sanchez:2006p331}. We performed
several tests using simulated and real spectra with a broad range of
intensities in order to assess the level of contamination that the extracted
spectra show due to the cross-talk. We tested the new method varying the
relative intensity of the spectra in the central and adjacent fibres, the
pixel extraction apertures and the average width of the Gaussian profiles. We
found that in the extreme cases (very bright adjacent fibres) the cross-talk
is suppressed by 95\%, being almost negligible for the range of spectral
intensities compared to the instrumental misalignments and distortions found
during conventional observations of the PINGS sample (see
\autoref{fig:cross}).\\

In summary, the two most important sources of error in the data reduction
arise from the sky subtraction and the flux calibration. 
We have made several tests and applied new techniques in each case in order
to minimise the magnitude of the uncertainties. 
We found that the sky subtraction cannot be applied as part of any standard
reduction pipeline and has to be considered on an individual basis, depending
on the observing mode, configuration and strategy. 
On the other hand,  the maximum expected error in the absolute flux
calibration for those objects in which a broad-band imaging recalibration was
applied is of the order of 20\%, being slightly larger in a narrow
blue spectral region $\lambda <$ 3800 \AA. For those objects without
broad-band imaging, the absolute flux calibration error is of the order of
30\%. However, the absolute error is better than 20\% for objects observed
during photometric conditions and/or with high S/N. The colors, spectral
features and gradients are completely reproduced when comparing our data with
previously  published long-slit observations (see \autoref{sec:integrated}).

DAR effects were found to be negligible due to the large size of the spatial
elements of PPAK, we did not find strong evidence of DAR in the several test
pointings analysed. However we flagged all individual IFU positions for which
the observations were performed with an airmass $>$ 1.3.
With respect to cross-talk, the new extraction method proved to
suppress this effect to a negligible level for the range of spectral
intensities in the PINGS observations.
Furthermore, no evidence of any second-order contamination has been found
during the data reduction and analysis of the PINGS data set. As explained in
\autoref{sec:obs}, the contamination of the 2nd order, up to 7200 \AA, is
expected to be lower than 1/10000, being negligible for our science case.

\section{The PINGS data}
\label{sec:data}

The PINGS data set contains more than 50000 flux calibrated spectra for a
sample of 17 galaxies covering in total an observed spectroscopic area of nearly 80
arcmin$^2$. \autoref{tab:observations} shows a summary of the observations,
including the number of individual IFU pointings observed for the mosaic of
each galaxy, the observational status of the mosaicking, the total number of
spectra, and additional individual comments on each object.
The spectroscopic data set samples the observed objects with fibre circular
apertures of $\sim$\,2.7 arcsec in diameter, covering the optical wavelength range
between $\sim$\,3700\,--\,7100 \AA, which includes the most prominent
recombination and collisionally excited emission lines from \oii
\lam\lam\,3727,\,3729 to \sii \lam\lam\,6717,\,6731.

It is beyond the scope of this article to extract all the information
potentially available in this huge spectroscopic data base. Independent
studies for individual objects, regions and galaxy subsamples will be presented
in a series of future papers. We consider here a few relevant scientific examples that
have been extracted from the data set to demonstrate the use of these data.
In this section 
\footnote{The analysis of this section was partly based on a set of IDL routines
created in order to handle and visualise the 2D spectroscopic
mosaics. Although they were first developed for the specific PINGS data
format, they can be implemented to any IFS data based on RSS files and
corresponding position tables. The PINGS software is freely available at
\url{http://www.ast.cam.ac.uk/research/pings/}.}
we present the integrated spectra and emission line
intensities for selected \hh regions of the IFS mosaics of NGC\,1058
and NGC\,3310, with a comparison to previously published data. Furthermore,
we obtained emission line maps for NGC\,1058 and we present a qualitative
description of the 2D distribution of physical properties inferred from the
line intensity maps. Finally, we present the spectra of the SN 2007gr observed
in NGC\,1058 as an example of the utility of wide-field IFS surveys to find
potential SNe progenitors and their environmental properties.

\subsection{Integrated spectra}
\label{sec:integrated}

\begin{figure*}
  \includegraphics[height=7cm]{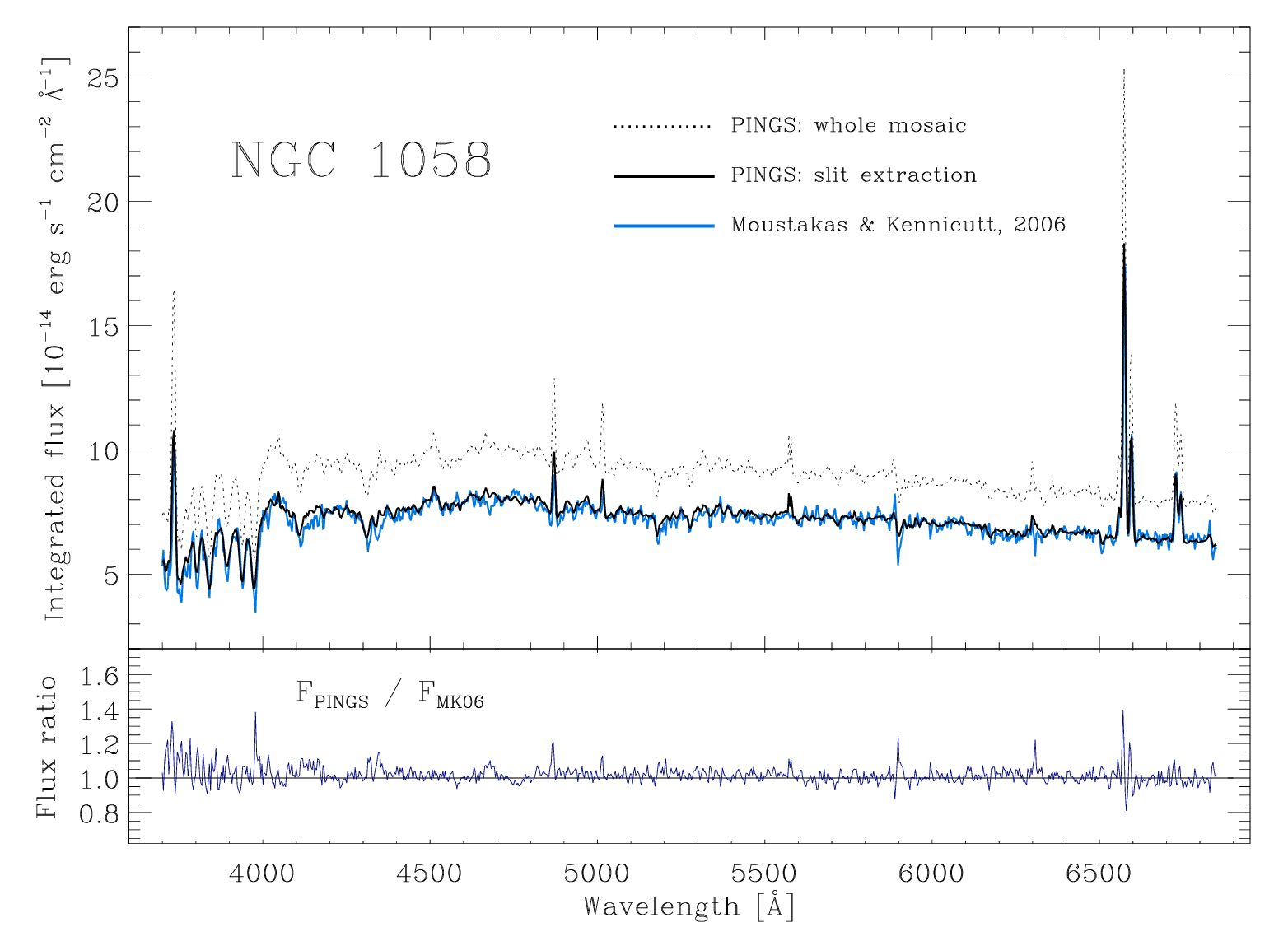}
  \includegraphics[height=7cm]{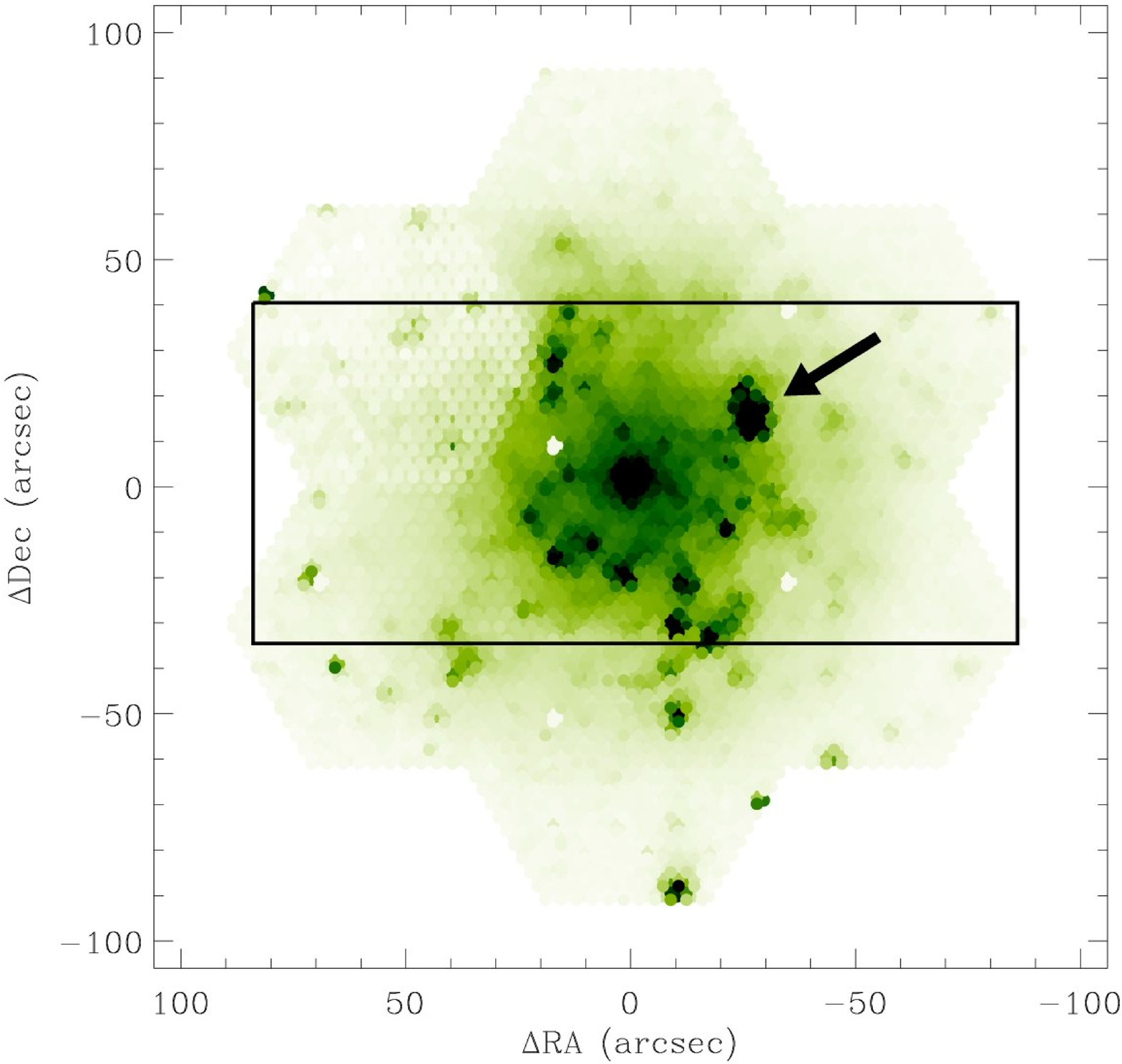}
  \includegraphics[height=7cm]{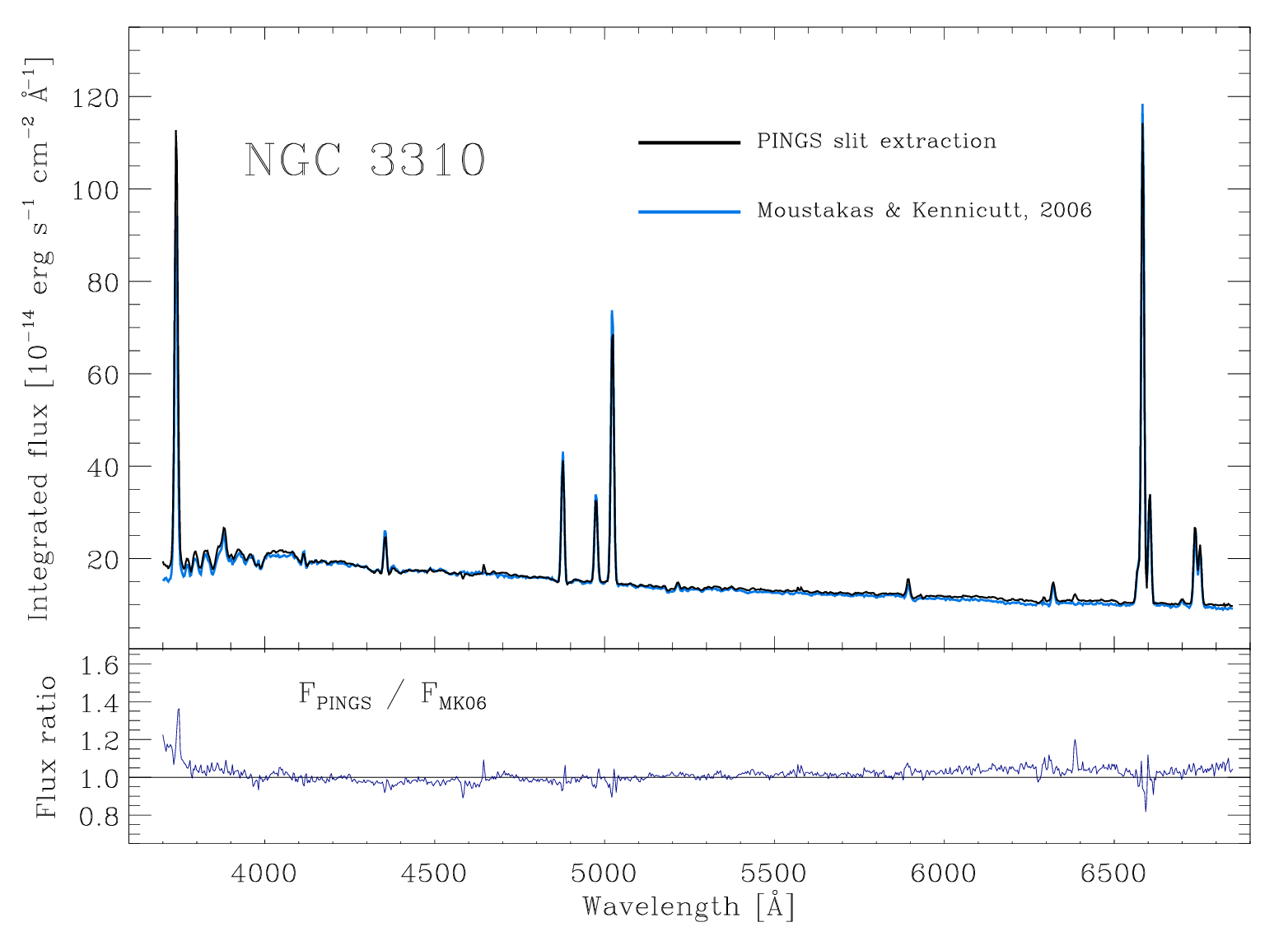}
  \includegraphics[height=7cm]{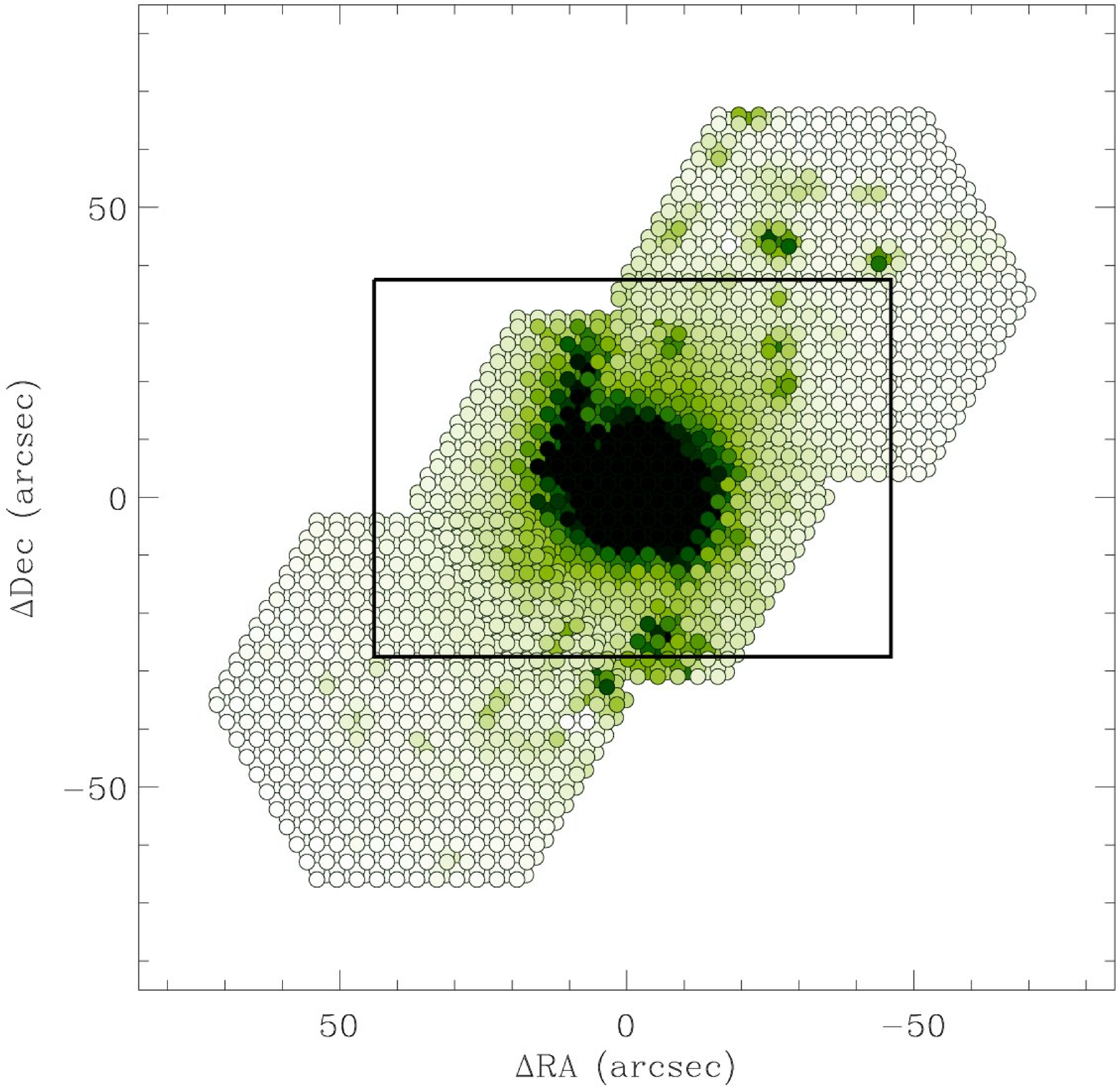}
  \caption{Top panels: Integrated spectra of NGC\,1058
  derived from the PINGS slit-extraction (thick solid line) and the whole
  mosaic (dotted line) spectra compared to the \citetalias{Moustakas:2006p307}
  data (blue line in the online version), as
  explained in the text. Flux in units of 10$^{-16}$ erg s$^{-1}$ cm$^{-2}$
  \AA$^{-1}$.
  Bottom panels: Comparison of the integrated spectra
  of NGC\,3310 with the same color coding as the figure above. For both
  galaxies, the diagrams on the right represent narrow-band intensity maps
  centred at H$\alpha$ showing the mosaic pattern and the rectangular
  slit-apertures of MK06. The arrow in the NGC\,1058 mosaic diagram shows the position of
  SN\,2007gr. The contours of the individual fibres have been drawn in the
  case of NGC\,3310.}
  \label{fig:integ}
\end{figure*}

The study of the optical integrated light provides a powerful tool to
investigate the physical properties of galaxies at different epochs in the
history of the universe. 
Spectral investigations based on integrated optical
spectrophotometry have been used to explore the main drivers of galaxy evolution,
e.g. star formation rate, star formation history, stellar mass, chemical
abundance, dust content, etc. \citep[e.g.][]{Kauffmann:2003p3500,Tremonti:2004p1138}.
The analysis of the integrated spectra in nearby objects can be used to assess
the limitations imposed by high-redshift surveys, such as their limiting
magnitude and incomplete spatial coverage (or aperture bias), factors that may
be important given that many physical properties of galaxies vary depending on
the geometry and position (e.g. stellar populations, metallicity, extinction,
etc.).
A solution to the incomplete spatial coverage of spectroscopic observations on
nearby objects consists on using sequential methods that use time to scan a
target while recording the spectral information.
The standard method for obtaining integrated
spectra in nearby objects was developed by \citet{Kennicutt:1992p2671}. The
so-called drift-scanning technique consists of a scan perpendicular to the slit
over the optical extent of the galaxy during a single exposure. Using this
method \citealt{Moustakas:2006p307} (hereafter MK06)
obtained spatially integrated optical (3600\,--\,6900 \AA) spectrophotometry
for 417 nearby galaxies of a diverse range of galaxy types, which were later
used to study several integrated galactic properties (SFR, metallicity, etc.).

One by-product of 2D spectroscopy and IFS data sets is the intrinsic
capability of adding up all the spectra within an observed field or mosaic
into a single spectrum, i.e. using the IFU as a large-aperture spectrograph to
obtain the integrated spectra of a given FOV.
The PINGS sample is an ideal data set for this purpose given that the
spectroscopic mosaics cover, in most cases, the entire optical radius of
the galaxy. The integrated spectra derived from PINGS can be used to study the
real average spectroscopic properties of a given nearby, large angular size
galaxy, as opposed to previous studies that attempted to describe their
average properties by the analysis of individual spectra taken from different
regions, or by targeting objects with a limited extraction aperture which
recovers only a fraction of the total optical light. 
In this section we present examples of high signal-to-noise integrated spectra
for NGC\,1058 and NGC\,3310 obtained by co-adding the spectra from their
corresponding mosaics using different simulated apertures applied to the
IFS mosaic. We compare these data with previously published integrated spectra
from \citetalias{Moustakas:2006p307}. In \citetalias{paperII}, we present the
integrated spectra of NGC\,628 with a more elaborated analysis of the
integrated stellar populations and nebular abundances.

\autoref{fig:integ} shows in the left upper panel the comparison of the
integrated spectrum of NGC\,1058 extracted from the PINGS mosaic to the
drift-scan spectrum obtained by \citetalias{Moustakas:2006p307} of this
galaxy, at the same spectral resolution. The right upper panel shows the PINGS
spectroscopic mosaic diagram of
NGC\,1058 with intensity levels corresponding to a ``narrow-band'' 100\,\AA\
width image extracted from the IFS data centred at H$\alpha$. The rectangular
shape shows the slit-aperture used in the drift-scan technique by
\citetalias{Moustakas:2006p307} to obtain the integrated spectrum of this
galaxy\footnote{ 
The simulated extraction slit was registered over the surface of the galaxy by
visual comparison to corresponding diagram shown by MK06, therefore there is
an uncertainty of the correct position of the extraction slit over the IFS
mosaic of the order of $\sim$ 2 arcsec.}.
Three foreground stars and the supernova SN\,2007gr (see
\autoref{sec:sn}) were removed from the co-added spectra (the arrow shows the
position of SN\,2007gr between two foreground stars). The black solid line
spectrum in the left upper panel was obtained by extracting from the PINGS
mosaic all the individual spectra within the area enclosed by the rectangular
shape and taking into account the overlapping regions and the covering
fraction of the fibres due to the dithering technique applied to this
galaxy.

The blue line corresponds to the spectrum obtained by
\citetalias{Moustakas:2006p307} using the drift-scan technique. The dotted
line was derived after co-adding all the spectra of the seven IFU positions
shown in \autoref{fig:integ} and considering the dithering overlaps. This
spectrum is constructed out of 6951 individual spectra covering the whole
optical $B_{25}$ mag arcsec$^{-2}$ radius of NGC\,1058. In this case we can see
clearly the effect of the incomplete spatial coverage when comparing the
integrated spectrum within the slit-aperture of
\citetalias{Moustakas:2006p307} and the whole optical radius covered by the
PINGS mosaic (dotted line). Although it is beyond the scope of this paper to
give a more quantitative comparison, we can notice from a purely qualitative
point of view that the determination of some physical properties may vary when
using both the slit-aperture and the whole mosaic integrated spectra, probably
by as much as $\sim$\,15\,--\,20\%.

In a similar way, the bottom panels of \autoref{fig:integ} show the integrated
spectra of the starburst galaxy NGC\,3310. The black line shows the integrated
spectrum obtained after co-adding all the spectra within the rectangular pattern
corresponding to the \citetalias{Moustakas:2006p307} slit-aperture and
considering all the overlaps in the mosaic due to the dithering observing
method and the non-standard construction of the mosaic. No foreground stars
were found within this field. The blue line corresponds to the
\citetalias{Moustakas:2006p307} drift-scan integrated spectrum. In this case
the PINGS extraction is spatially incomplete when compared to the rectangular
area of the \citetalias{Moustakas:2006p307} aperture. Nevertheless, all the
spectral features are reproduced quite well when we compare the PINGS spectrum
with the \citetalias{Moustakas:2006p307} data.
The apparent disagreement seen in some of the emission lines can be
explained by the spatial incompleteness of the IFS mosaic with respect to the
MK06 slit, and the presence of a bright \hh region at
($\Delta\alpha$,\,$\Delta\delta$) $\sim$ (--10,--30), which is located at the
edge of the extraction slit and presents particularly strong emission in the
\oiii lines. A small shift of the reported position of the MK06 slit results
in a slightly different integrated spectrum due to this strong \hh region,
causing that the peaks of the emission lines differ in height.
A stronger effect is seen in the blue spectral region ($\lambda <$\,3800 \AA)
which is expected due to the intrinsic problems of the spectrophotometric
calibration in this regions as explained in \autoref{sec:errors}.
The integrated spectrum of NGC\,3310 using all the
fibres in the three dithered positions differs very little from the extracted
one, and it is not shown for the sake of clarity.

The spectroscopic mosaics of NGC\,1058 and NGC\,3310 were not re-calibrated
using the broad-band imaging technique described in \citetalias{paperII}, and
therefore they are not calibrated for the absolute zero point
flux. Nevertheless, both the spectral shape and the spectral features of the
PINGS extracted spectra corresponding to the slit-aperture of
\citetalias{Moustakas:2006p307} match remarkably well with these previously
published, spectrophotometrically calibrated data, showing the quality of
the PINGS IFS data set.

\subsection{Emission line maps}
\label{sec:emission}

\begin{figure*}
  \includegraphics[height=7cm]{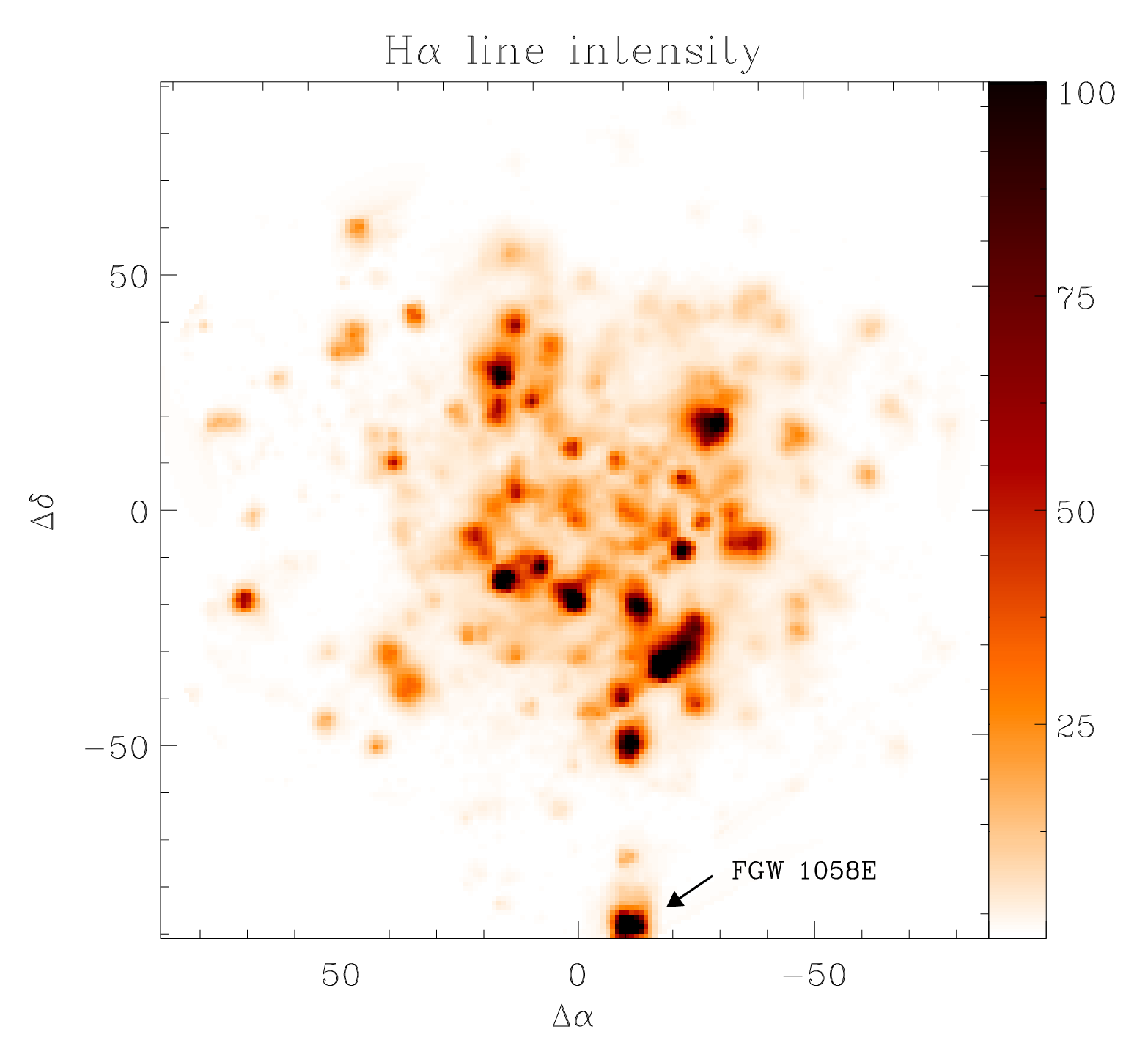}
  \includegraphics[height=7cm]{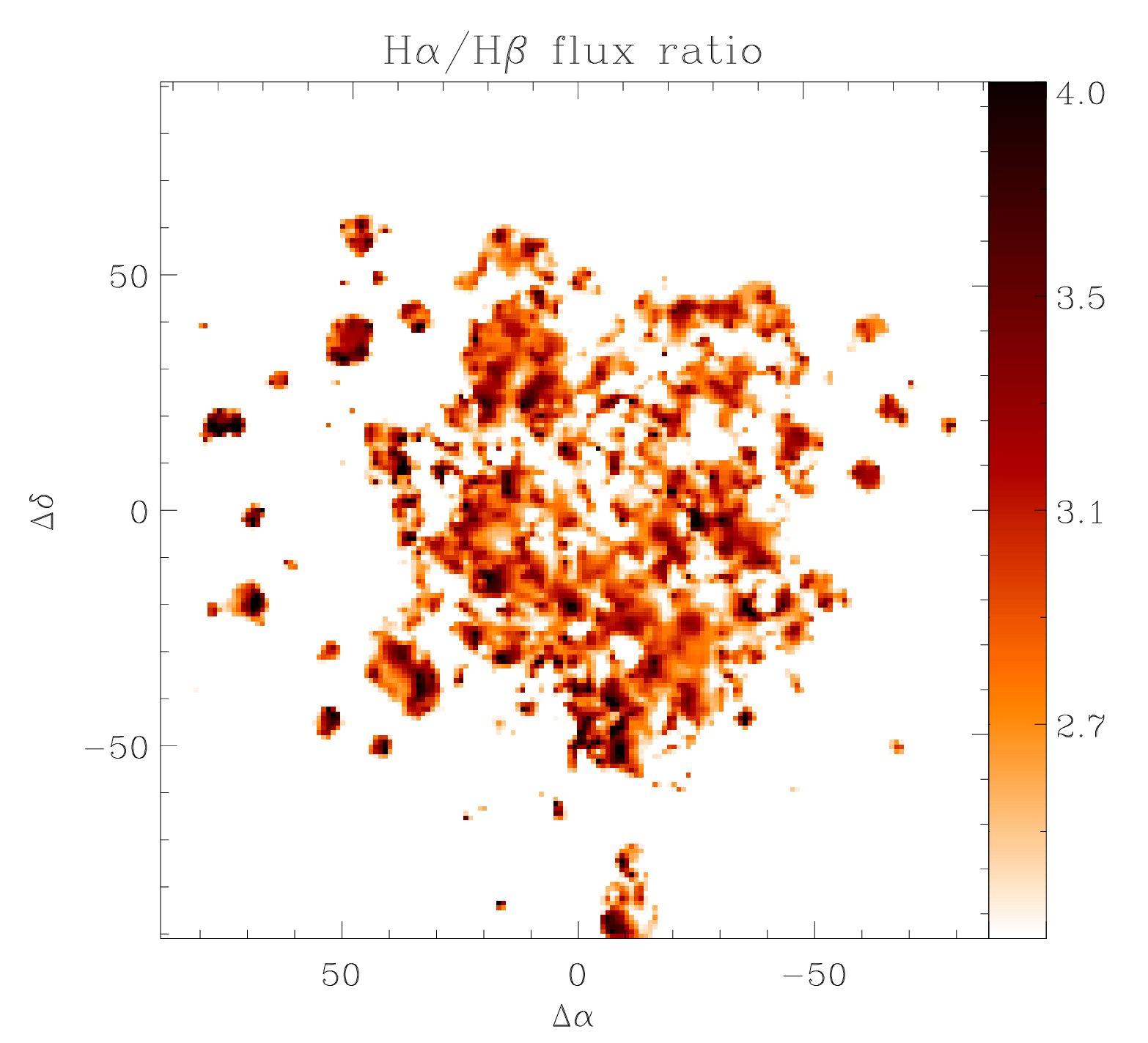}
  \includegraphics[height=7cm]{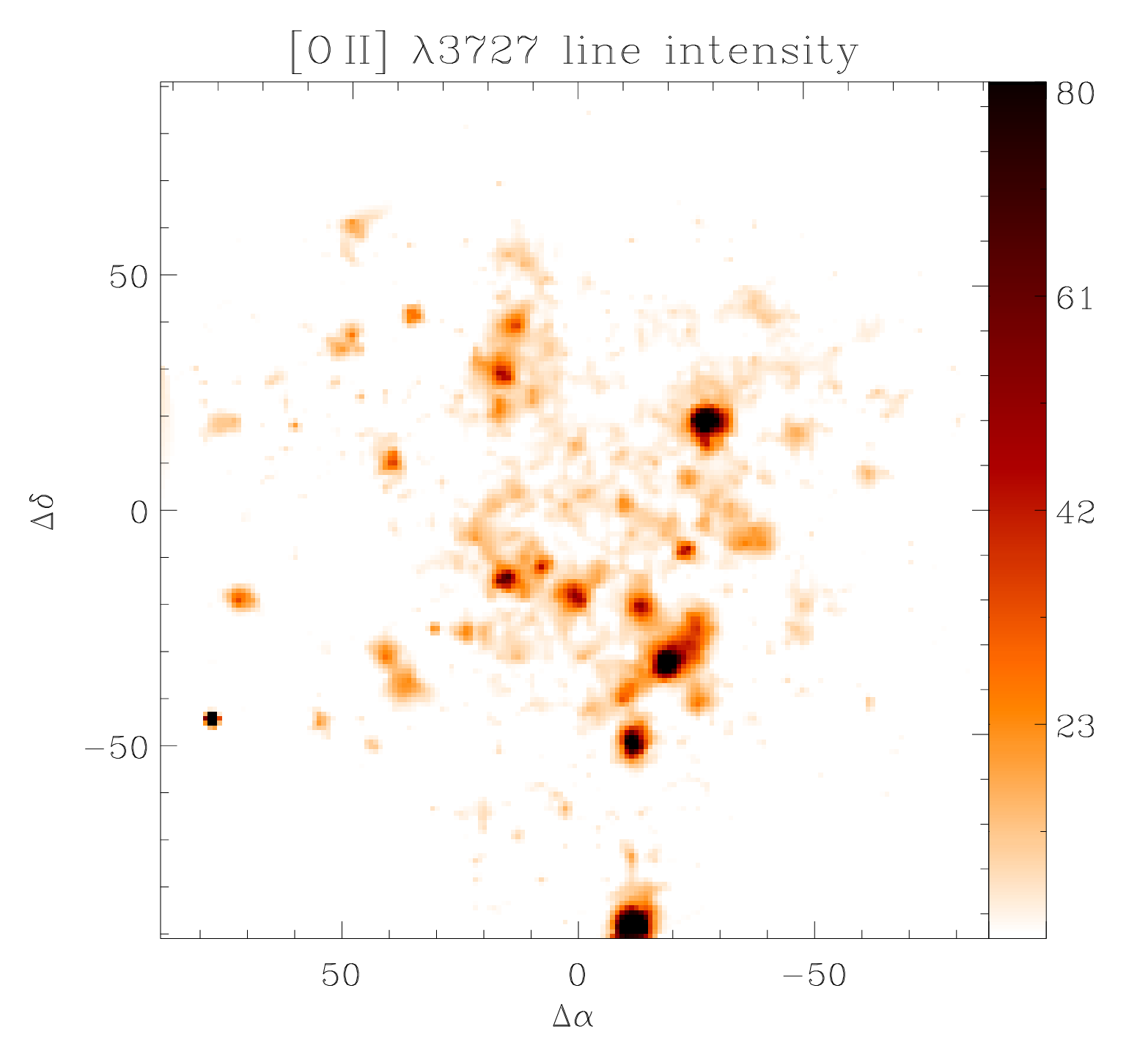}
  \includegraphics[height=7cm]{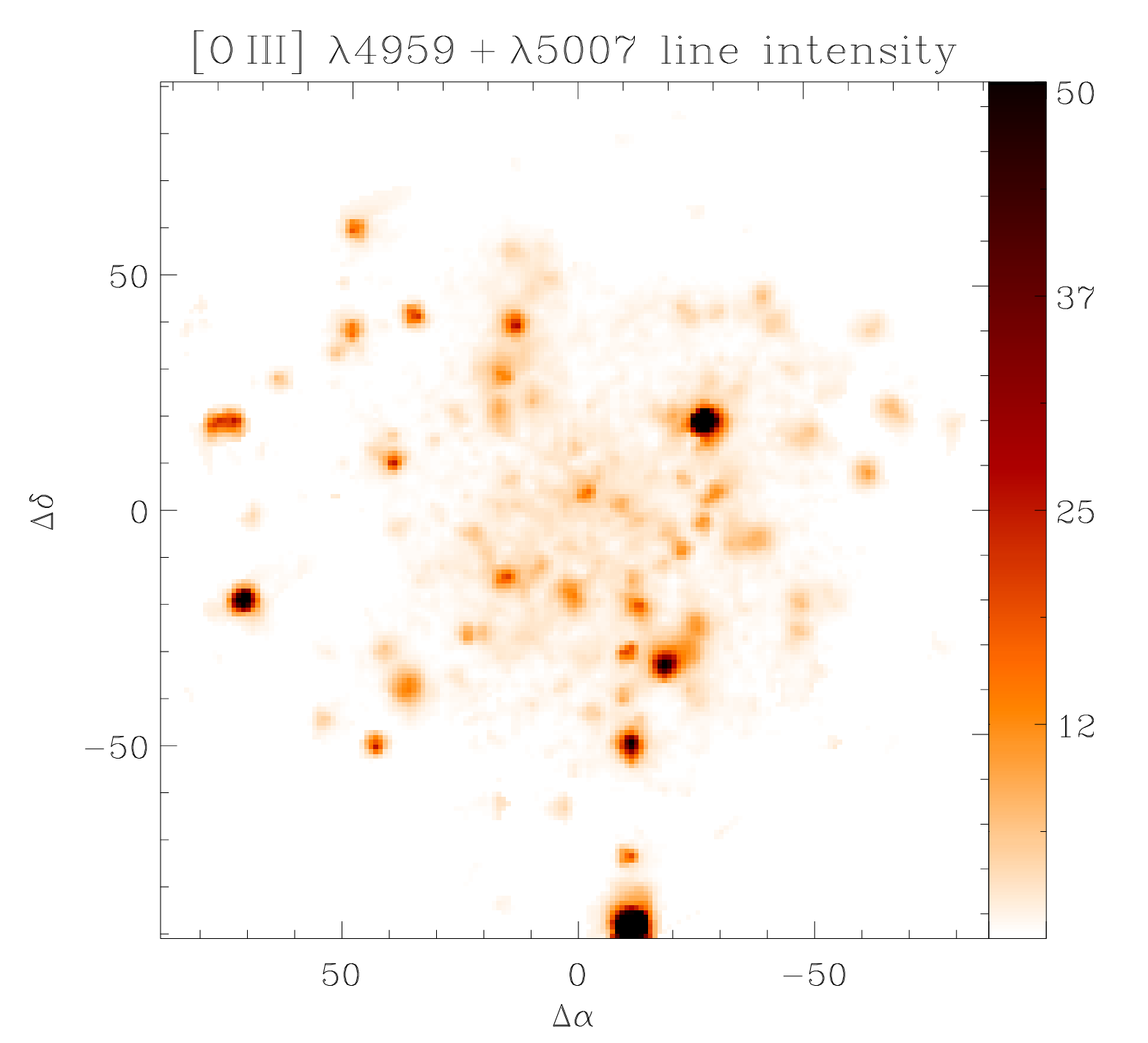}
  \caption{Emission line maps obtained from the IFS mosaic of NGC\,1058
  (without the outlying pointing). In the top-left panel, H$\alpha$ line
  intensity map, top-right H$\alpha$/H$\beta$ flux line ratio. Bottom-left,
  [O\,{\small II}] \lam3727, bottom-right [O\,{\small III}]
  \lam4959\,+\,\lam5007 intensity maps. 
  All maps were obtained by fitting a single Gaussian function to the emission
  line for each single spectrum in the mosaic. The intensity map is then
  reconstructed by interpolating the recovered flux at the location of each
  spectrum as described in the text. All maps are in units of 10$^{-16}$ erg
  s$^{-1}$ cm$^{-2}$ arcsec$^{-2}$. No correction for dust extinction was
  applied to any map.}
  \label{fig:emission}
\end{figure*}

One of the main objectives of the PINGS project is to obtain complete
maps of the emission-line intensities which could then be analysed to describe
the 2D spatially-resolved distribution of the physical properties of the
sample. The PINGS spectroscopic mosaics allow us to obtain for the
first time a complete 2D view in the optical wavelength range of the main
emission lines used in typical abundance diagnostics methods, and important
spectral features useful for the analysis of the underlying stellar
populations. 

The ionized gas in spiral galaxies exhibits a complex structure associated
morphologically with star-forming regions located mainly along the spiral
arms. Previous attempts to perform a 2D analysis have made use of narrow-band
and Fabry-Perot imaging at different spectral widths. However, in some
cases the narrow-band imaging includes the contribution of more than one
single emission line, such as the case of the H$\alpha$ imaging, which
includes the \nii $\lambda\lambda$\,6548,\,6584 doublet, or the \sii
density sensitive doublet at $\lambda\lambda$\,6717,\,6731. This factor limits
the utilization of these techniques to study just the basic parameters of the
ionized gas, under the assumption of fixed line ratios. The great advantage of
IFS arises from the fact that we are able to deblend emission lines at any
discrete spatial location, and to ultimately produce maps of individual
emission lines.

In order to extract any physical information from the data set, we need first
to identify the detected emission lines of the ionized gas and to decouple
their emission from the stellar population in each individual spectrum of the
IFS mosaic.
We used population synthesis to model and subtract the stellar continuum
underlying the nebular emission lines. This technique results in emission-line
measurements corrected (to a first-order) for stellar absorption. The details of
this process are described in \citetalias{paperII}, however we present briefly
here the scheme followed to decouple the stellar population and the emission
lines in our database: i) A set of emission lines is identified from any strong \hh
region of the galaxy. ii) For each spectrum in the data set, the underlying
stellar population is fitted by a linear combination of a grid of Single
Stellar Populations (SSP), masking all the nebular and sky emission lines. The
template models are selected in order to cover the widest possible range of
ages and metallicities. We consider the effects of dust extinction by varying
A$_V$ from 0 to 1 mag at $\Delta$\,0.2 mag. iii) We subtract the fit stellar
population from the original spectrum to get a residual pure emission-line
spectrum. iv) Finally, we derive the intensities for each detected emission
line.

Individual emission-line fluxes were measured in each spectrum by considering
spectral window regions of $\sim$ 200\,\AA. 
We performed a simultaneous multi-component fitting using a single Gaussian function (for
each emission line contained within each window) plus a low order polynomial
(to describe the local continuum and to simplify the fitting procedure) using
FIT3D \citep{Sanchez:2006p3300}.
The central redshifted wavelengths of the emission lines were fixed and since
the FWHM is dominated by the spectral resolution, the widths of all the lines
were set equal to the width of the brightest line in this spectral region. This
procedure decreases the number of free parameters and increases the accuracy
of the deblending process (when required).
Line intensity fluxes were then measured by integrating the observed intensity of
each line. The statistical uncertainty in the measurement of the line flux was
calculated by propagating the error associated to the multi-component fitting
and considering the signal-to-noise of the spectral region.

We applied this line-fitting method to NGC\,1058 and
created emission line maps by interpolating the intensities derived for each
individual line in each individual spectrum, correcting for the dithering
overlapping effects. The interpolation was performed using E3D, adopting a
natural-neighbour, non-linear interpolation scheme, with a final scale
of 1''/pixel in the resulting maps. The data at the location of bright
foreground stars in the field were masked prior to any interpolation, in order
to decrease the effects of their contamination.

As a prime example of this technique for the purpose of this paper, we
calculated the following emission line maps: H$\alpha$, H$\beta$, the
doublet-blended \oii \lam3727, and \oiii
$\lambda\lambda$\,4959,\,5007. In order to prevent contamination by low
signal-to-noise data, we masked all pixels that correspond to an integrated
flux per fibre below 10$^{-16}$ erg s$^{-1}$ cm$^{-2}$.
\autoref{fig:emission} shows the emission line maps in units of 10$^{-16}$ erg
s$^{-1}$ cm$^{-2}$ arcsec$^{-2}$. The top panels
show the distribution of the star-forming regions in the galaxy traced by the
H$\alpha$ intensity map (left) and the distribution of dust extinction, outlined by the
H$\alpha$/H$\beta$ flux ratio (right). From these maps we note that,
although many of the most intense emission regions are located along the inner
spiral arms of the galaxy, some of them lie in outer regions not associated
with any spiral structure (e.g. to the south of the galaxy, at $\Delta$RA $\sim$\,--10,
$\Delta$Dec $\sim$\,--90 arcsec). 
On the other hand, the H$\alpha$/H$\beta$ flux ratio shows a smooth
distribution, especially concentrated in the inner part of the galaxy with
strong peaks in the outlying star-forming regions.
As a matter of fact, the very intense outlying regions of star-formation are
responsible for the difference found in between the integrated spectra of NGC\,1058
when comparing the central slit-aperture of \citetalias{Moustakas:2006p307} to
the whole mosaic (see \autoref{sec:integrated}). NGC\,1058 is known to have a larger
than average H\,I-to-optical size, and deep H$\alpha$ imaging has revealed the
existence of recent massive star formation out to and beyond two optical radii
(\citealt{Ferguson:1998p224}, hereafter FGW98).
However, the small number of \hh regions
analysed in previous studies of NGC\,1058 has prevented a complete explanation
of the existence of these extreme outer regions, and of the intrinsic scatter
of the abundance gradient seen in this galaxy. 
Given the spatial information provided by the PINGS spectroscopic
mosaic, we could in principle study the properties of the
outer \hh regions with those of the inner disk and test the predictions of
chemical evolution models and the behavior of the chemical abundances not only
as a function of increasing radius, but taking into consideration the
morphology of the galaxy.

The bottom panels show the line intensity distribution of \oii
\lam3727 (left), and \oiii \lam4959\,+\,\lam5007
(right). For both cases, we notice strong emission regions at $\Delta$RA
$\sim$ --20 arcsec, and again several outlying emission
regions that do not follow the spiral pattern of the galaxy. More definite
interpretations than the qualitative ones described here can be achieved with
a full spectroscopic analysis of the dust content, ionization conditions and
metallicity distribution of the the whole IFS mosaic.

\begin{table*}
\centering
\label{tab:fluxes}
\caption[Line ratios vs. literature]{Comparison of the PINGS emission line
  intensities for two \hh regions in NGC\,1058 and one in NGC\,3310 with
  previous observations and published line ratios. For NGC\,1058 we compare
  with the regions E and H by \citet{Ferguson:1998p224}, and the Jumbo \hh
  region in NGC\,3310 observed by \citet{Pastoriza:1993p3323}. For NGC\,1058,
  the coordinates shown represent approximate offsets of the \hh regions with
  respect to the galaxy nucleus in the PINGS mosaic in the format
  ($\Delta$RA,~$\Delta$Dec) with NE positive. All values for NGC\,1058 are the
  observed line ratios. In the case of the NGC\,3310 Jumbo \hh region, the
  observed and redenning corrected ratios for PINGS are shown as Pas\,93 only
  shows corrected line intensities (see text for details). All line ratios are normalised to
  H$\beta$. *\,Observed H$\beta$ flux in units of 10$^{-16}$
  erg~s$^{-1}$\,cm$^{-2}$.}



{\scriptsize
\begin{tabular*}{\textwidth}{@{\extracolsep{\fill}} lrrrrrrrr }

\hline
\\[-4pt]

& \multicolumn{4}{c}{\footnotesize\sc NGC\,1058} & & \multicolumn{3}{c}{\footnotesize\sc NGC\,3310 Jumbo H~{\tiny II} region} \\[2pt]

\cline{2-5} \cline{7-9} \\[-6pt]

& \multicolumn{2}{c}{(--10.6,~--87.9)} & \multicolumn{2}{c}{(140.6,~90.0)}  && \multicolumn{2}{c}{PINGS} & \multicolumn{1}{c}{Pas\,93} \\[2pt]

\multicolumn{1}{c}{Line} & \multicolumn{1}{c}{PINGS} & \multicolumn{1}{c}{FGW 1058E} & \multicolumn{1}{c}{PINGS} & \multicolumn{1}{c}{FGW 1058H} &&
                           \multicolumn{1}{c}{F($\lambda$)/F(H$\beta$)} & \multicolumn{1}{c}{I($\lambda$)/I(H$\beta$)} & 
                           \multicolumn{1}{c}{I($\lambda$)/I(H$\beta$)} \\  [2pt]

\hline
\\[-6pt]

[O{\tiny~II}]\,~$\lambda$3727    &   3.033~$\pm$~0.262  & 2.739~$\pm$~0.064   &   2.261~$\pm$~0.130 &   1.950~$\pm$~0.052  &&   2.388~$\pm$~0.132  &   3.028~$\pm$~0.352 &  2.59~$\pm$~0.040    \\[2pt]
[Ne{\tiny~III}]\,~$\lambda$3869  &   0.170~$\pm$~0.018  &                     &   0.188~$\pm$~0.026 &                      &&   0.180~$\pm$~0.020  &   0.223~$\pm$~0.048 &  0.18~$\pm$~0.040    \\[2pt]
H{\tiny8}~+~He{\tiny~I}\,~$\lambda$3889
                                         &   0.164~$\pm$~0.018  &                     &   0.111~$\pm$~0.021 &                      &&   0.126~$\pm$~0.021  &   0.156~$\pm$~0.044 &  0.13~$\pm$~0.039    \\[2pt]
H$\epsilon$\,~$\lambda$3970              &   0.162~$\pm$~0.018  &                     &   0.173~$\pm$~0.025 &                      &&   0.115~$\pm$~0.020  &   0.140~$\pm$~0.043 &  0.17~$\pm$~0.053    \\[2pt]
H$\delta$\,~$\lambda$4101                &   0.209~$\pm$~0.022  &                     &   0.228~$\pm$~0.029 &                      &&   0.184~$\pm$~0.020  &   0.218~$\pm$~0.039 &  \multicolumn{1}{c}{0.21:}    \\[2pt]
H$\gamma$\,~$\lambda$4340                &   0.386~$\pm$~0.039  &                     &   0.447~$\pm$~0.046 &                      &&   0.406~$\pm$~0.018  &   0.455~$\pm$~0.046 &  \multicolumn{1}{c}{0.40:}    \\[2pt]
[O{\tiny~III}]\,~$\lambda$4363   &   \multicolumn{1}{c}{\ldots} &             &   \multicolumn{1}{c}{\ldots} &             &&   0.017~$\pm$~0.004  &   0.019~$\pm$~0.006 &  0.018~$\pm$~0.005  \\[2pt]
He{\tiny~I}\,~$\lambda$4471      &   0.029~$\pm$~0.004  &                     &   0.052~$\pm$~0.012 &                      &&   0.034~$\pm$~0.004  &   0.037~$\pm$~0.009 &  0.036~$\pm$~0.009  \\[2pt]
H$\beta$\,~$\lambda$4861                 &   1.000~$\pm$~0.050  & 1.000~$\pm$~0.033   &   1.000~$\pm$~0.051 &   1.000~$\pm$~0.037  &&   1.000~$\pm$~0.038  &   1.000~$\pm$~0.082 &  1.000~$\pm$~0.053  \\[2pt]
[O{\tiny~III}]\,~$\lambda$4959   &   0.745~$\pm$~0.038  & 0.786~$\pm$~0.026   &   1.137~$\pm$~0.057 &   1.148~$\pm$~0.042  &&   0.936~$\pm$~0.048  &   0.919~$\pm$~0.067 &  0.740~$\pm$~0.007   \\[2pt]
[O{\tiny~III}]\,~$\lambda$5007   &   2.301~$\pm$~0.115  & 2.298~$\pm$~0.076   &   3.227~$\pm$~0.163 &   3.282~$\pm$~0.117  &&   2.857~$\pm$~0.060  &   2.779~$\pm$~0.143 &  2.310~$\pm$~0.009   \\[2pt]
He{\tiny~I}\,~$\lambda$5876      &   0.109~$\pm$~0.011  &                     &   0.223~$\pm$~0.024 &                      &&   0.133~$\pm$~0.011  &   0.114~$\pm$~0.013 &  0.099~$\pm$~0.005  \\[2pt]
[O{\tiny~I}]\,~$\lambda$6300     &   0.050~$\pm$~0.006  &                     &   0.069~$\pm$~0.009 &                      &&   0.060~$\pm$~0.008  &   0.050~$\pm$~0.010 &  0.051~$\pm$~0.004  \\[2pt]
[S{\tiny~III}]\,~$\lambda$6312   &   \multicolumn{1}{c}{\ldots}  &            &  \multicolumn{1}{c}{\ldots}  &             &&   0.011~$\pm$~0.004  &   0.009~$\pm$~0.006 &  0.011~$\pm$~0.003  \\[2pt]
[O{\tiny~I}]\,~$\lambda$6363     &   \multicolumn{1}{c}{\ldots}  &            &  \multicolumn{1}{c}{\ldots}  &             &&   0.021~$\pm$~0.006  &   0.017~$\pm$~0.008 &  0.014~$\pm$~0.005  \\[2pt]
[N{\tiny~II}]\,~$\lambda$6548    &   0.142~$\pm$~0.023  & 0.135~$\pm$~0.005   &   0.065~$\pm$~0.013 &   0.099~$\pm$~0.009  &&   0.189~$\pm$~0.018  &   0.152~$\pm$~0.020 &  0.150~$\pm$~0.005   \\[2pt]
H$\alpha$\,~$\lambda$6563                &   3.114~$\pm$~0.156  & 3.231~$\pm$~0.106   &   3.135~$\pm$~0.158 &   2.980~$\pm$~0.106  &&   3.574~$\pm$~0.056  &   2.870~$\pm$~0.179 &  2.880~$\pm$~0.004   \\[2pt]
[N{\tiny~II}]\,~$\lambda$6584    &   0.413~$\pm$~0.045  & 0.414~$\pm$~0.014   &   0.187~$\pm$~0.022 &   0.179~$\pm$~0.009  &&   0.548~$\pm$~0.038  &   0.439~$\pm$~0.055 &  0.460~$\pm$~0.003   \\[2pt]
He{\tiny~I}\,~$\lambda$6678      &   0.037~$\pm$~0.004  &                     &   0.020~$\pm$~0.006 &                      &&   0.045~$\pm$~0.006  &   0.035~$\pm$~0.008 &  0.032~$\pm$~0.003  \\[2pt]
[S{\tiny~II}]\,~$\lambda$6717    &   0.373~$\pm$~0.028  & 0.362~$\pm$~0.012   &   0.190~$\pm$~0.020 &   0.238~$\pm$~0.011  &&   0.315~$\pm$~0.014  &   0.249~$\pm$~0.032 &  0.270~$\pm$~0.006   \\[2pt]
[S{\tiny~II}]\,~$\lambda$6731    &   0.270~$\pm$~0.023  & 0.257~$\pm$~0.009   &   0.134~$\pm$~0.014 &   0.148~$\pm$~0.008  &&   0.248~$\pm$~0.015  &   0.196~$\pm$~0.025 &  0.220~$\pm$~0.006   \\[3pt]

\hline
\\[-5pt]

I($\lambda$5007)/I($\lambda$4959)        &   3.09~$\pm$~0.15    &  2.92~$\pm$~0.14    &    2.83~$\pm$~0.20  &   2.85~$\pm$~0.15    &&   3.05~$\pm$~0.05  &   3.02~$\pm$~0.08 & 3.12~$\pm$~0.03         \\[3pt]
*~F(H$\beta$)\,\,$\lambda$4861             &  \multicolumn{1}{c}{236.2}  & \multicolumn{1}{c}{210.5}  &  \multicolumn{1}{c}{11.6}   &   \multicolumn{1}{c}{19.5} && \multicolumn{1}{c}{1096}  && \multicolumn{1}{c}{1690} \\[3pt]

\hline
\end{tabular*}
}
\end{table*}

%
%
%

%
%
%

\subsection{Emission line ratios: comparison with literature}
\label{sec:lit}

Several galaxies in the PINGS sample have been spectroscopically studied
previously by different authors. As a consistency test of the quality of our
data we performed a comparison of several emission line ratios as measured by the
procedure described in the previous section with selected \hh regions from the
literature for which the authors published the emission line ratios for
different species. \autoref{tab:fluxes} shows a comparison of the PINGS emission line
intensities for three of those regions: FGW\,1058E and FGW 1058H, analysed 
by \citetalias{Ferguson:1998p224}, and the Jumbo \hh region in NGC\,3310 observed by
\citealt{Pastoriza:1993p3323}, (hereafter Pas93).

\citetalias{Ferguson:1998p224} observed a total of 8 \hh regions in NGC\,1058,
however, regions FGW\,1058A to D are located in the inner part of the galaxy
and their identification is somewhat unclear (see Figure 2 from
\citetalias{Ferguson:1998p224}). Regions FGW\,1058F and FGW\,1058G
fall outside the observed FOV of PINGS. On the other hand, FGW\,1058E is a
bright \hh region located at $\sim$ (--10,~--88) arcsec in
($\Delta$RA,~$\Delta$Dec) units with respect to the galaxy centre
(see \autoref{fig:emission}), while FGW\,1058H is an outlying \hh region
located at $\sim$ (140,~90) in the PINGS mosaic (see \autoref{fig:maps_1}).
For this comparison we selected these two objects as both fall within the FOV
observed by PINGS and are uniquely distinguishable from the H$\alpha$ maps/images.

In the case of FGW\,1058, we used for this comparison a 4 arcsec circular
aperture centered on the fibre with the strongest emission in H$\alpha$,
assuming that the long-slit observation was placed in this region as
\citetalias{Ferguson:1998p224} did not give details of the observation of each
specific object and their given offsets are only approximate with respect to
the centre of the galaxy (even the extraction aperture is uncertain as they
only quote a size range from 3 to 15 arcsec depending on the seeing and on the
size of the object in question).
In the case of FGW\,1058H, the identification was relatively simple as this is a
small and well-defined outlying \hh region. The emission line ratios for this
region were obtained from a single fibre at the quoted
position. \citetalias{Ferguson:1998p224} published the observed and reddening corrected
emission line intensities for 9 spectral lines, in \autoref{tab:fluxes} we
compare our results with the observed un-corrected emission line fluxes
only. Additional spectral lines and atomic species detected and measured in
the PINGS data are also included in \autoref{tab:fluxes}.

\citetalias{Pastoriza:1993p3323} performed an optical and near-IR spectroscopic
analysis of circumnuclear \hh regions (at less than 400 pc from the nucleus)
and the Jumbo \hh region in NGC\,3310. The identification of the circumnuclear
\hh regions is somewhat difficult in the PINGS mosaic given the linear scale of
the regions and the size of the fibres. Therefore we chose to analyse the
bright Jumbo \hh region for which emission line intensities were measured by
\citetalias{Pastoriza:1993p3323}. The slit position, aperture and PA are well
described by these authors, however we did not choose to simulate an aperture
in the PINGS mosaic as the aperture of the slit used by
\citetalias{Pastoriza:1993p3323} (1.5 arcsec) is smaller than the size of a
single PPAK fibre. Instead, we chose the fibre within this region with the strongest
emission in H$\alpha$, which corresponds to an offset (--10.6,~--2.2) in
($\Delta$RA,~$\Delta$Dec) with respect to the galaxy centre in the PINGS
mosaic.

\citetalias{Pastoriza:1993p3323} quoted only the reddening corrected line
intensities, therefore for this comparison, \autoref{tab:fluxes} shows the
observed and reddening corrected line ratios obtained from the PINGS spectrum
extracted for this region. 
The PINGS observed line intensities listed in \autoref{tab:fluxes} were corrected
for reddening using the Balmer decrement according to the reddening function of
\citet{Cardelli:1989p136}, assuming $R \equiv A_V/E(B-V) = 3.1$. Theoretical
values for the intrinsic Balmer line ratios were
taken from \citet{Osterbrock:2006p2331}, assuming case B recombination, an
electron density of $n_e = 100$ cm$^{-3}$ and an electron temperature $T_e =
10^4$ K. We have used only the H$\alpha$/H$\beta$ ratio to deduce the
logarithmic reddening constant c(H$\beta$), obtaining a value of 0.32 for the
PINGS spectrum, whereas \citetalias{Pastoriza:1993p3323} obtained c(H$\beta$)
= 0.29 for the same region. We did not correct the emission line fluxes for
underlying stellar absorptions. Formal errors in the derived line ratios were
estimated by summing in quadrature the uncertainty in flux calibration, the
statistical error in the measurement of the line flux and the error in the
c(H$\beta$) term.

Despite the ambiguity due to the exact location and extraction aperture of the
observed spectra, from \autoref{tab:fluxes} we can see that there is a very
good agreement between the previously published emission line ratios and the
PINGS observations for these three \hh regions. The strongest deviation is
found in the \oii \lam3727 as we expected, since this line falls in the
spectral region at which the instrumental low sensitivity increases the flux
calibration error. The comparison with the NGC\,1058 regions is more
straightforward as these are the observed values without any further
correction. The comparison of the emission intensities with the NGC\,3310 Jumbo
\hh region has to be made carefully since the values of
\citetalias{Pastoriza:1993p3323} were corrected for underlying absorption and
using a different reddening curve. Nevertheless, the reddening corrected
values of PINGS are in good agreement with the values derived by
\citetalias{Pastoriza:1993p3323}, we even detected and measured correctly the
very faint \oiii \lam4363 line, showing the good quality of the sky
subtraction despite the presence of the Hg \lam4358 sky line. The line
strength of the \oiii \lam4363 line combined with the
redshift of NGC\,3310 makes it fall in the region of ``detectability'' as
described in the simulation presented in \autoref{sec:simu}.

An additional assessment of the quality of the data is given by the value of
the I(\lam5007)/I(\lam4959) ratio which can be predicted from atomic
theory and observed in high signal-to-noise astronomical spectra.
According to current atomic computations, the theoretical value for this
intensity ratio is 2.98 \citep{Storey:2000p3365}.
In \autoref{tab:fluxes} we show this ratio for the three \hh regions.
The observed central values for the region
FGW\,1058E in both datasets are in good agreement with the theoretical value. In the case of
FGW\,1058H both values are quite similar but differ within the errors from the
theoretical value. For the NGC\,3310 Jumbo \hh region, the central value
observed by PINGS is closer to the theoretical value for both uncorrected and
corrected line ratios compared to the \citetalias{Pastoriza:1993p3323} value.

\begin{figure}
  \includegraphics[width=0.49\textwidth]{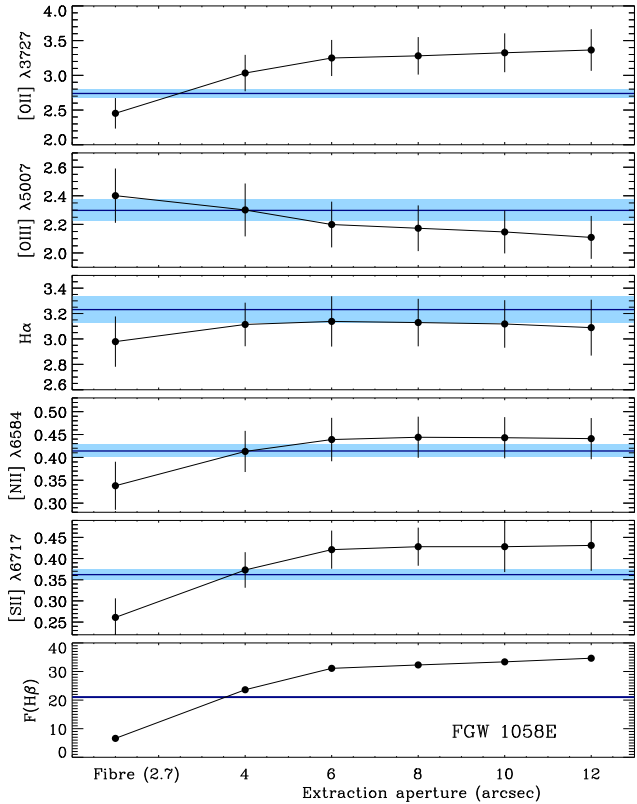}
  \caption[Aperture extraction ratios]{Variation of the emission line ratios as
    a function of integration aperture of the \hh region FGW 1058E. The
   horizontal line/band in each panel shows the value/error derived by
   \citetalias{Ferguson:1998p224}. All emission line intensities are
   normalized to H$\beta$. The observed integrated fluxes of H$\beta$ are in units of
   10$^{-15}$ erg s$^{-1}$ cm$^{-2}$.}
  \label{fig:apertures}
\end{figure}

The flux observed by PINGS in the H$\beta$ line for FGW\,1058E is slightly higher
than the one measured by \citetalias{Ferguson:1998p224}, contrary to the case
of FGW\,1058H, where the flux in PINGS for the same line is somehow smaller,
reflecting the unknown aperture extraction for the
\citetalias{Ferguson:1998p224} long-slit spectrum. In the case of the
NGC\,3310 Jumbo \hh region, the flux measured by PINGS is somewhat smaller
than the one measured by \citetalias{Pastoriza:1993p3323}.

The 2D character of the PINGS data allows us to study the variation of the
spectra within a given area that would be otherwise taken as a single \hh region.
In \autoref{fig:apertures} we show the effect of the extraction aperture on
the emission line intensities for the \hh region
FGW\,1058E. \citetalias{Ferguson:1998p224} considered FGW\,1058E as a single
\hh region, however a closer look using the dithered spectroscopic mosaic shows that this
region is actually a complex composed of several knots and substructures with
varying emission fluxes in the most prominent lines. 
In order to examine the difference in the emission line ratios in this region,
we take as a central position the fibre with the strongest emission in
H$\alpha$ within this area, with an integration aperture of 2.7 arcsec diameter. We then
take concentric circular apertures of different sizes (ranging from 4 to 12
arcsec in diameter), we integrate the spectra within these apertures to obtain 
spectra from which we measure a different set of emission line intensities.

The five top panels of \autoref{fig:apertures} show the variation of the
emission line ratios obtained at different extraction aperture sizes for some
relevant lines (normalised to H$\beta$). The second data point corresponds to the value shown in
\autoref{tab:fluxes} (4 arcsec aperture), the horizontal line in each panel shows the central
value and the error bar obtained by \citetalias{Ferguson:1998p224} (the dark/light
blue colour line/band in the online version). The bottom panel shows the
integrated flux of H$\beta$ at each aperture. From \autoref{fig:apertures} we
can see that the emission line ratios measured using different extraction
apertures vary considerably as a function of the aperture size, and that in
most cases the dispersion of the central values is larger than the error of
the measurements, reflecting that this is a physical effect. All emission line
ratios tend to converge to a certain value as the aperture size
increases. Note that at the flux level in H$\beta$ measured by
\citetalias{Ferguson:1998p224} for this region, all the emission line
intensity ratios, as measured by PINGS, are basically the same (within errors)
as the values derived by \citetalias{Ferguson:1998p224}.

From this exercise we note that, to a first-order, the emission line ratios
measured for a given \hh region may significantly depend on the
morphology of the region, on the slit (fibre) position, on the extraction
aperture and on the signal-to-noise of the observed spectrum. All these
effects should be taken into account when deriving physical quantities from
spectroscopic studies of \hh regions.

\begin{figure}
  \includegraphics[width=0.49\textwidth]{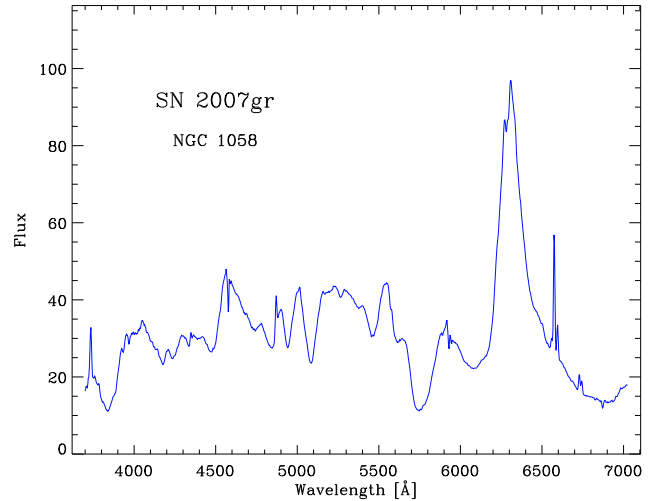}
  \caption[SN 2007gr in NGC\,1058]{Extracted spectrum of the SN 2007gr in the
    mosaic of NGC\,1058. The flux units are 10$^{-16}$ erg s$^{-1}$ cm$^2$
    \AA$^{-1}$. Some nebular emission lines are seen superimposed to the SN
    spectrum.}
  \label{fig:sn}
\end{figure}

\subsection{SN 2007\lowercase{gr} in NGC 1058}
\label{sec:sn}

As another example of the potential of the PINGS spectroscopic data, we present the
spectrum of a particular object observed during the mosaicking of the galaxy
NGC\,1058. The observation of this galaxy was carried out during the nights of
the 7th and 8th of December 2007. During this period we were able to observe
the spectrum of the Supernova 2007gr, previously discovered by the Lick
Observatory Supernova Search on the 15th of August 2007
\citep{Chornock:2007p2679}.

SN\,2007gr was located at 24.8'' West and 15.8'' North of the nucleus of NGC
1058, with coordinates RA 02$^{\rm h}$\,43$^{\rm m}$\,27.$^{\rm s}$98 and Dec:
+37$^{\circ}$\,20$^{\rm m}$\,44.$^{\rm s}$7.
In the mosaic diagram of NGC\,1058 shown in \autoref{fig:integ}, the arrow
shows the location of SN\,2007gr between two bright foreground
stars. \autoref{fig:sn} shows the optical spectrum of the supernova extracted
from the IFS mosaic, showing the typical spectrum of a Type Ic core collapse
supernova, with the lack of hydrogen, helium and silicon absorption
lines. Emission lines of \oii \lam3727, H$\alpha$, \nii
\lam\lam\,6548,\,6584 and \sii \lam\lam\,6717,\,6731 are clearly
seen on top of the SN spectra, probably reflecting the environment of the \hh
region in which this supernova exploded.

As described by \citet{Valenti:2008p2680}, SN 2007gr showed an average peak
luminosity but unusually narrow spectral lines and an almost flat photospheric
velocity profile.  SN 2007gr motivated an extensive observational campaign for
several reasons: it was discovered at a very early stage (5 days after the
explosion), it was located in a relatively close distance galaxy, it was the
nearest stripped-envelope carbon-rich SNe ever observed, and a suitable
candidate for progenitor search \citep{Valenti:2008p2680}. 
Wide-field IFS may become one of
the main resources for SNe research groups to find SNe progenitors in
previously observed galaxies. The progenitor supergiant stars ($M >$ 8
$M_{\odot}$) that at the end of their lives explode as core-collapse
supernovae may show a strong stellar spectrum that could be recovered from 2D
spectroscopic maps with enough spatial resolution. Furthermore, the nebular
emission from the spatially adjacent spectra could provide information on the
environment in which these SNe explosions occur.

\section{Summary}
\label{sec:summary}

We have presented the PPAK IFS Nearby Galaxies Survey: PINGS, a project
designed to construct 2D spectroscopic mosaics of 17 nearby disk galaxies in
the optical wavelength range. This project represents one of the first
attempts to obtain continuous coverage spectra of the whole surface of a
galaxy in the nearby universe.

The final sample includes different galaxy types, including normal, lopsided,
interacting and barred spirals with a good range of galactic properties and SF
environments with multi-wavelength public data. 
The spectroscopic data set comprises more than 50000 individual spectra, covering an
observed area of nearly 80 arcmin$^2$.  The data set will be supplemented with
broad band and narrow band imaging for those objects without publicly available
images in order to maximise the scientific and archival value of the observations.
Future plans consider to release the PINGS data set in order to make it freely
available to the scientific community as a PPAK legacy project.

The primary scientific objectives of PINGS are to
obtain emission-line maps and moderate-resolution 
spectra of the underlying stellar population across the disks of the
galaxies. These spectral maps will allow us to study the spatial
distribution of the physical properties of the ionized gas and the stellar
components in galaxies, solving the limitations imposed by the small FOV and
spectral coverage of previous attempts at obtaining the 2D information of a galaxy.
PINGS will provide the most detailed knowledge of star formation, gas
chemistry and the variations across a late-type galaxy. This information is
very relevant for interpreting the integrated colours and spectra of high
redshift sources. The details provided by PINGS will contribute significantly
to the study of the chemical abundances and the global properties of galaxies.

We have assessed very carefully all sources of errors and uncertainties found
during the intrinsically complex reduction of IFS observations. This complexity
is further deepened if one considers building an IFU spectroscopic mosaic of
a given object for which the observations were performed at very different
stages, for some objects even spanning a period of years.

To demonstrate the use of these data, we presented a few relevant scientific
examples that have been extracted from the data set. 
We presented the integrated spectra of the IFS mosaics of NGC\,1058 and
NGC\,3310 and a comparison with previously published data. Furthermore,
we obtained emission line maps for NGC\,1058 and we presented a qualitative
description of the 2D distribution of physical properties inferred from the
line intensity maps. 
We performed a comparison of the emission line ratios obtained for a number of
selected \hh regions from the literature, we showed that the emission line
intensities obtained from the PINGS data agree with previously published
studies, and that the line ratios derived from a single \hh region may depend
strongly on the position and aperture size of the slit/fibre (among other
factors) for a given spectroscopic study.
Finally, we showed the spectra of the SN 2007gr observed
in NGC\,1058 as an example of the utility of wide-field IFS surveys to find
potential SNe progenitors and their environmental properties.
Independent studies for individual objects, regions and galaxies samples will
be presented in future papers.
As a highlight of this project, the spectroscopic mosaicking of one of the
sample objects: NGC\,628, represents the largest area ever covered by an IFU
observation. In \citetalias{paperII} we present a complete analysis of the
integrated and spatially resolved properties of this galaxy.

\section*{Acknowledgments}

FFRO would like to acknowledge the Mexican National Council for Science and
Technology (CONACYT), the Direcci{\'o}n General de Relaciones Internacionales
(SEP), Trinity College, the Cambridge Philosophical Society, and the Royal
Astronomical Society for the financial support to carry out this research
SFS and AD thank the Spanish Plan Nacional de Astronom{\'i}a programmes
AYA\,2005-09413-C02-02 and AYA\,2007-67965-C03-03 respectively.  SFS
acknowledges the Plan Andaluz de Investigaci{\'o}n of Junta de Andaluc{\'i}a
as research group FQM\,360.








\bibliographystyle{mn2e}
\bibliography{astroph}

\label{lastpage}

\end{document}